\documentclass{memoir}

\usepackage{hyperref}

\usepackage{amsmath} 
\usepackage{gensymb}
\usepackage{listings}

\usepackage{color}
\usepackage{xcolor}
\usepackage{latexsym}
\usepackage{framed}
\usepackage{multirow}  
\usepackage{pdflscape} 
\usepackage{longtable}
\usepackage{array}
\usepackage{wrapfig}
\usepackage{palatino}
\usepackage{graphicx}

\usepackage{calc}

\usepackage[T1]{fontenc}

\usepackage[font={footnotesize,sf},labelfont=bf]{caption}
\definecolor{navyblue}{rgb}{0,0,0.55}     
\definecolor{darkgreen}{rgb}{0,0.3,0.3}
\definecolor{commentgreen}{rgb}{0,0.45,0}
\definecolor{keywordscolor}{rgb}{0.1,0,0.5}
\definecolor{verylightgray}{gray}{0.95}   

\hypersetup{
  colorlinks=true,
  linkcolor=navyblue,
  urlcolor=navyblue,
  pdfnewwindow=true,
  citecolor=darkgreen}

\setsecnumdepth{subsection} 
\abnormalparskip{0.75em}

\newcolumntype{L}[1]{>{\raggedright\arraybackslash}p{#1}}
\newcolumntype{R}[1]{>{\raggedleft\arraybackslash}p{#1}}

\newcommand*{\mytitlepage}{
  \begingroup
  \vfill
  \hbox{
    \hspace*{0.05\textwidth}
    \rule{2pt}{0.95\textheight}
    \hspace*{0.05\textwidth}
    \parbox[b]{0.85\textwidth}{
      \vbox{
          {\noindent\HUGE\bfseries Dissecting the \\[0.5\baselineskip]
          NVidia Turing T4 GPU \\[0.5\baselineskip]
          via Microbenchmarking}\\[4.5\baselineskip]

          \normalfont
          {\Large Technical Report }\\[4\baselineskip]

          {\Large Zhe Jia\\[0.35\baselineskip]
          Marco Maggioni\\[0.35\baselineskip]
          Jeffrey Smith\\[0.35\baselineskip]
          Daniele Paolo Scarpazza
          }\par

          \vspace{0.22\textheight}
          \includegraphics[trim=12.5cm 13.65cm 13.5cm 13.25cm, clip=true, totalheight=0.8cm]{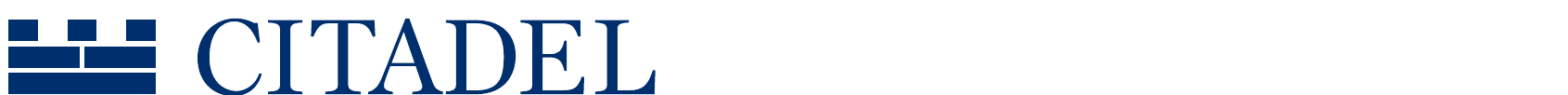}
          \\[0.4em]
          {High Performance Computing R\&D Team\\[0.1em]
          Citadel, 131~S.~Dearborn St., Chicago, Ill. 60603.
          }\par
          \vspace{0.03\textheight}
      }
    }
  }
  \vfill
  \null
  \endgroup
  }

\begin{document}
\thispagestyle{empty}
\newpage
\mytitlepage
\sloppy
\thispagestyle{empty}\newpage
{
  \small
  \noindent
  Copyright \copyright \ 2019, Citadel Enterprise Americas, LLC. All
  rights reserved.

  \noindent
  Chicago, United States of America.

  \noindent
  Citadel grants arXiv.org a perpetual, non-exclusive license to
  distribute this manuscript.

  \noindent
  This presentation solely reflects the analyses and views of the authors.
  No recipient should interpret this presentation to represent the
  general views of Citadel or its personnel.  Facts, analyses, and
  views presented herein have not been reviewed by, and may not
  reflect information known to other Citadel professionals.

  \noindent
  This report strives to be objective, neutral and impartial.  The
  authors did not receive any form of compensation or reimbursement in
  conjunction with this work, and are not employed by NVidia. All
  results in this report are meant to be reproducible by any
  researcher who recreates the experimental setup described by
  the authors, except for good-faith mistakes. No such mistakes are
  known to the authors at the time of publication. Findings are solely
  based only on the authors' experiments and understanding,
  independent of NVidia's confirmation of methods or data.

  \noindent
  The authors welcome all feedback.  The corresponding author is
  Daniele P. Scarpazza. Send feedback preferably to his e-mail address:
  \href{mailto:daniele.scarpazza@citadel.com}{daniele.scarpazza@citadel.com}.

  \vspace{1ex}
  \noindent
  \makebox[\textwidth]{\centering
    \begin{tabular}{ll}
      \toprule
      Edition  & Date \\
      \midrule
      First    &  March 19, 2019 \\
      \bottomrule
    \end{tabular}
  }
  \vspace{1ex}

  \noindent
  The authors make this edition, and will make future editions, of
  this document available on the \href{http://arXiv.org}{arXiv.org}
  e-print service owned and operated by Cornell University.
  The authors presented material contained in this report at the 2019
  \href{http://www.gputechconf.com}{GPU Technology Conference}, March
  18-21, 2019, San Jose, California.

  \noindent
  All product names, trademarks and registered trademarks are property
  of their respective owners. All company, product and service names
  mentioned are for identification purposes only. Use of
  these names, trademarks and brands does not imply endorsement.

}

\newpage
\setcounter{page}{3}

\chapter*{Summary}
In 2019, the rapid rate at which GPU manufacturers refresh their
designs, coupled with their reluctance to disclose microarchitectural
details, is still a hurdle for those software designers who want to
extract the highest possible performance from GPUs.

Last year, these very reasons motivated us to dissect the Volta GPU
architecture using microbenchmarks. We presented our
findings at NVidia's GPU Technology Conference
(GTC2018)~\cite{zhe2018gtc} and published them in a technical
report~\cite{zhe2018}.

The introduction in August 2018 of Turing~\cite{tu104}, NVidia's
latest architecture, pressed us to update our study. In this report,
we examine Turing and compare it quantitatively against previous
NVidia GPU generations. Specifically, we study the T4 GPU: a
low-power, small form-factor board aiming at inference applications.
We describe its improvements against its inference-oriented
predecessor: the P4 GPU based on the Pascal architecture. Both T4 and
P4 GPUs achieve significantly higher frequency-per-Watt figures than
their full-size counterparts.

We study the performance of the T4's Tensor Cores, finding a much
higher throughput on low-precision operands than on the P4 GPU. We
reveal that Turing introduces new instructions that express matrix
math more succinctly. We map Turing's instruction space, finding the
same encoding as Volta, and additional instructions. We reveal that the
Turing TU104 chip has the same memory hierarchy depth as the Volta
GV100; cache levels sizes on the TU104 are frequently twice as large
as those found on the Pascal GP104. We benchmark each constituent of
the T4 memory hierarchy and find substantial overall performance
improvements over its P4 predecessor. We studied how clock throttling
affects compute-intensive workloads that hit power or thermal limits.

Many of our findings are novel, published here for the first time. All
of them can guide high-performance software developers get closer to
the GPU's peak performance, as we illustrate with examples.

\chapter{Low-level details make a difference}

In this section, we use a practical example to motivate our claim that
a deep understanding of the architecture can help developers achieve
substantial speed-ups in certain, specific scenarios, although at the
cost of significant development effort.

It takes disproportionate effort to optimize code on the basis of a
deep understanding of its target architecture. This approach
frequently resorts to writing inline PTX assembly and, when pushed to
its extreme, to patching binary code in the pursuit of
specific SASS assembly that the compiler won't emit, following
undocumented instruction encoding formats, without any support from
NVidia's toolchain.

Whether the gains are worth the effort is a central question
but, ultimately, one that only you can answer depending on your
unique circumstances and pressure for performance. For a large
majority of GPU software developers, \emph{the answer is no}:
\begin {itemize}
  \item if one of the mature libraries provided by NVidia (such as
    cuBlas, cuFFT, cuSparse, cuRand, cuDNN, etc.) covers the
    computation desired, the performance obtained is close to ideal in
    most circumstances;
  \item in other applications, for which developers write CUDA code,
    NVCC usually emits efficient machine code, if the source code is
    written sufficiently well.
\end {itemize}

Rare are the cases where the pressure for performance justifies
extreme, low-level optimization. A prolific line
of research has traditionally focused on understanding GPU instruction
encoding~\cite{asfermi,askepler,nervana} precisely to improve the
performance of compute kernels~\cite{wong2010,zhang2017,mei2017}.  Our
prior work on Volta~\cite{zhe2018} also offered such an example: we
patched compiler-emitted code so that it used the register cache
better, and achieved a 15\% higher floating-point arithmetic
throughput.

\begin{figure}[t]
  \center
  \includegraphics[width=\columnwidth]{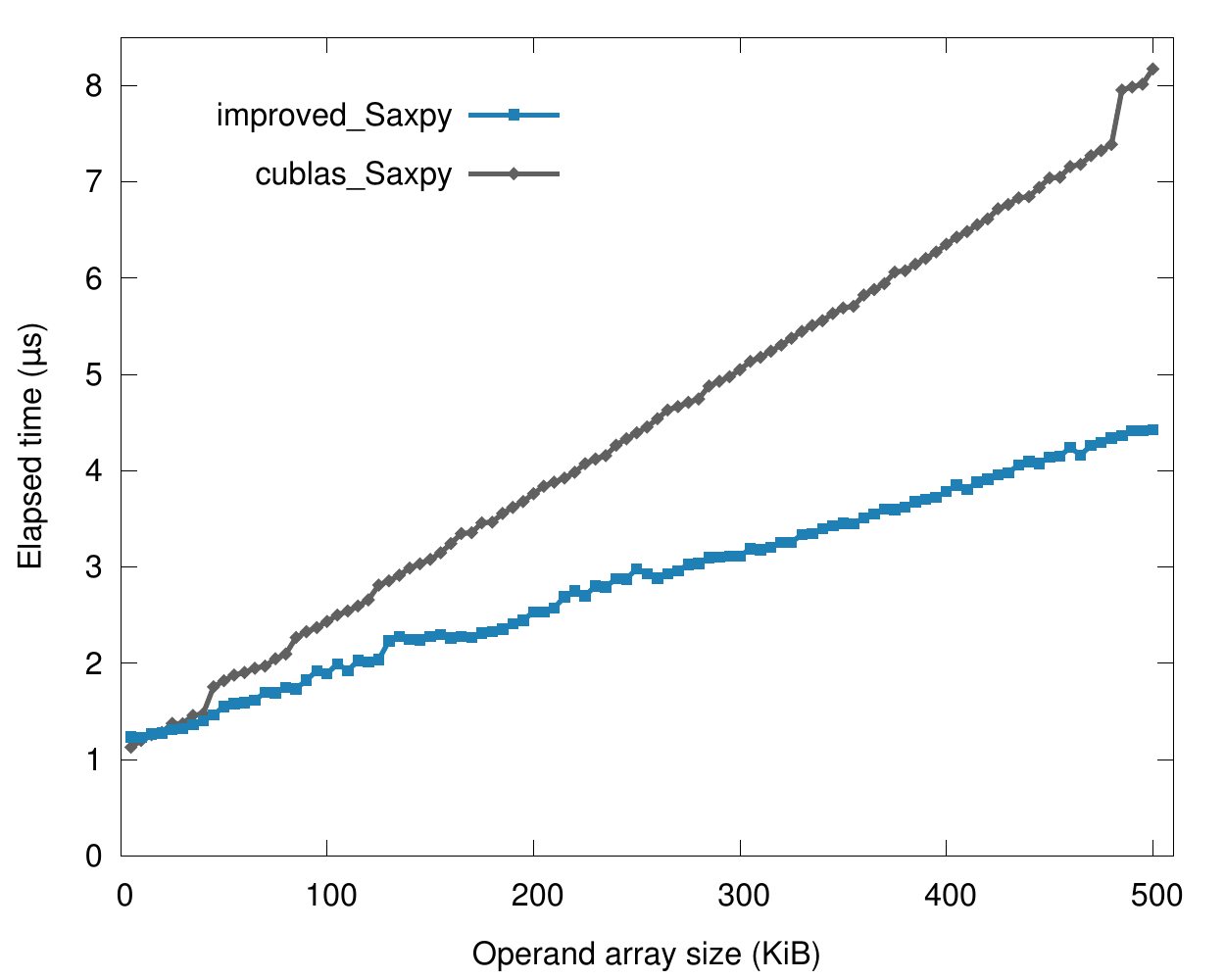}
  \caption{Performance of our improved \texttt{saxpy} implementation
    that uses 128-bit wide global memory access instructions, compared
    with \texttt{cublasSaxpy} from NVidia's cuBlas library, that uses
    32- and 64-bit wide instructions.  Elapsed time in microseconds;
    lower is better.}
  \label{fig:axpy_performance}
\end{figure}

This time, we show that the knowledge of Turing's instructions allows
designers to improve the performance of a common linear algebra
function (i.e., BLAS \texttt{?axpy}), whose library implementation for
single-precision operands contains memory access operations limited to
a 64-bit width. We show a simple replacement that uses vectorized
128-bit accesses, and improves performance substantially (see
Fig.~\ref{fig:axpy_performance}).

The \texttt{?axpy} workload we chose performs a scaled, element-wise
vector-vector sum, i.e.:
$$ \vec{y} := a \cdot \vec{x} + \vec{y} $$ where $a$ is the scale
factor. The workload is quite obviously of low arithmetic intensity,
which makes it memory bound. This means that any efficiency gain in
its access to global memory translates into a direct overall speedup.

The most straightforward way to compute this workload in C on a CPU
code is invoking BLAS function \texttt{cblas\_?axpy} (where
`\texttt{?}'  is a placeholder for the operand type, i.e., \texttt{s},
\texttt{d}, \texttt{c}, or \texttt{z}, respectively, for single- or
double-precision, real or complex operands).  In CUDA, the
corresponding function is \texttt{cublas?axpy}, from NVidia's cuBlas
library. For simplicity, we only focus on its single-precision variant
\texttt{cublasSaxpy}.

In our analysis, function \texttt{cublasSaxpy} calls, in turn,
\texttt{axpy\_kernel\_val} which, in its single-precisions
implementations shipped with the CUDA SDK version 10.0, contain load
and store instructions (from/to global memory) no wider than 64
bits. An excerpt of one such implementation follows:
\begin{lstlisting}[basicstyle={\scriptsize\ttfamily}]
Function : void axpy_kernel_val<float2, float2, 0>(
                        cublasAxpyParamsVal<float2, float2, float2>)
        .headerflags    @"EF_CUDA_SM75 EF_CUDA_PTX_SM(EF_CUDA_SM75)"
...
        /*01f0*/        LDG.E.64.SYS R2, [R2] ;
        /*0200*/        LDG.E.64.SYS R8, [R4] ;
...
        /*0330*/        LDG.E.64.SYS R2, [R2] ;
        /*0340*/        LDG.E.64.SYS R8, [R4] ;
...
\end{lstlisting}

\noindent This access width is suboptimal on Turing, especially
considering that:
\begin{enumerate}
\item Turing has only half as many load/store units per SM as Volta
  (16 vs. 32, as per public NVidia information~\cite{gv100,tu104});
\item a T4 device supports only half as many threads per SM than a
  V100 device (1,024 vs. 2,048); it is therefore harder to saturate the
  available memory bandwidth on Turing by just increasing block
  count.
\end{enumerate}

\noindent As a consequence, loading wider words per instruction is an effective
strategy to increase memory access throughput. We do so in our improved
implementation of the \texttt{Saxpy} kernel, that uses 128-bit vectorized
memory access instructions:

\begin{lstlisting}[basicstyle={\scriptsize\ttfamily}]
__global__ void improved_Saxpy( float *d_y, const float *d_x,
                                const float alpha, const uint32_t arraySize)
{
  // every thread process 4 elements at a time
  uint32_t tid = (threadIdx.x+blockIdx.x*blockDim.x)*4;
  // the elements that all threads on GPU can process at a time
  uint32_t dim = gridDim.x*blockDim.x*4;

  for(uint32_t i = tid; i < arraySize; i += dim)
    asm volatile ("{\t\n"
        // registers to store input operands
        ".reg .f32 a1,b1,c1,d1;\n\t"
        ".reg .f32 a2,b2,c2,d2;\n\t"

        // loading with vectorized, 128-bit instructions
        "ld.global.v4.f32 {a1,b1,c1,d1},[%0];\n\t"
        "ld.global.v4.f32 {a2,b2,c2,d2},[%1];\n\t"

        // core math operations
        "fma.rn.f32  a2,a1,%2,a2;\n\t"
        "fma.rn.f32  b2,b1,%2,b2;\n\t"
        "fma.rn.f32  c2,c1,%2,c2;\n\t"
        "fma.rn.f32  d2,d1,%2,d2;\n\t"

        // storing results with a vectorized, 128-bit write instruction
        "st.global.v4.f32 [%1],{a2,b2,c2,d2};\n\t"
        "}" :: "l"(d_x+i),"l"(d_y+i), "f"(alpha) : "memory"
    );
}
\end{lstlisting}

\noindent The PTX assembly that you see inlined in our code above
visibly uses 128-bit wide load and store instructions
\texttt{ld.global.v4.f32} and \texttt{st.global.v4.f32}, capable of
transferring a vector of four single-precision floating-point values
at a time.  (For simplicity and brevity, our implementation neglects
arrays whose size is not a multiple of 4.)

Inspection of the corresponding SASS code emitted by NVCC
confirms that global memory instructions are 128 bits wide:

\begin{lstlisting}[basicstyle={\scriptsize\ttfamily}]
    .headerflags    @"EF_CUDA_SM75 EF_CUDA_PTX_SM(EF_CUDA_SM75)"
    ...
    /*00d0*/         LDG.E.128.SYS R8, [R8] ;
    /*00e0*/         LDG.E.128.SYS R4, [R2] ;
    ...
    /*0110*/         FFMA R4, R8, c[0x0][0x170], R4 ;
    /*0120*/         FFMA R5, R9, c[0x0][0x170], R5 ;
    /*0130*/         FFMA R6, R10, c[0x0][0x170], R6 ;
    /*0140*/         FFMA R7, R11, c[0x0][0x170], R7 ;
    /*0150*/         STG.E.128.SYS [R2], R4 ;
    ...
\end{lstlisting}

The performance of this \texttt{improved\_Saxpy} code proves to be
significantly higher than the performance of \texttt{cublasSaxpy} (see
Figure~\ref{fig:axpy_performance}) except for trivially small arrays
($<$20 Kib), and it asymptotically tends to be almost twice as fast
for large ones.

In summary, with this example we demonstrated optimization
opportunities that are only accessible to a software designer who
possesses in-depth knowledge of Turing's instruction set and an
architectural-level understanding of its performance behavior: these
goals are the very subjects of this report.

\chapter{How Turing encodes instructions}

By systematically disassembling machine code that we hand-crafted and
that we sampled from representative CUDA libraries, using
\texttt{cuobjdump} and \texttt{nvdisasm}, we discovered the
instruction encoding formats adopted across the different GPU
architectures. Turing adopts the same format as Volta, which differs
from that of Pascal and Maxwell which, in turn, is different from
Kepler's, as we detail in this chapter.

Turing and Volta use 128 bits to encode both an instruction and its
associated scheduling control information\footnote{By control
  information in this context, we mean instruction scheduling
  decisions taken by the compiler, that the architecture must enforce;
  the next section discusses this topic in detail.}.  This is a
substantial departure from previous NVidia GPU architectures that used
one word per instruction (64-bit) to encode pure instruction
information, plus a separate 64-bit word every few instructions, to
encode control information associated those instructions.

The following example illustrates a Turing/Volta instruction, together
with its control information, as decoded by
\texttt{nvdisasm}~\cite{cuobj}:
\includegraphics[width=\columnwidth]{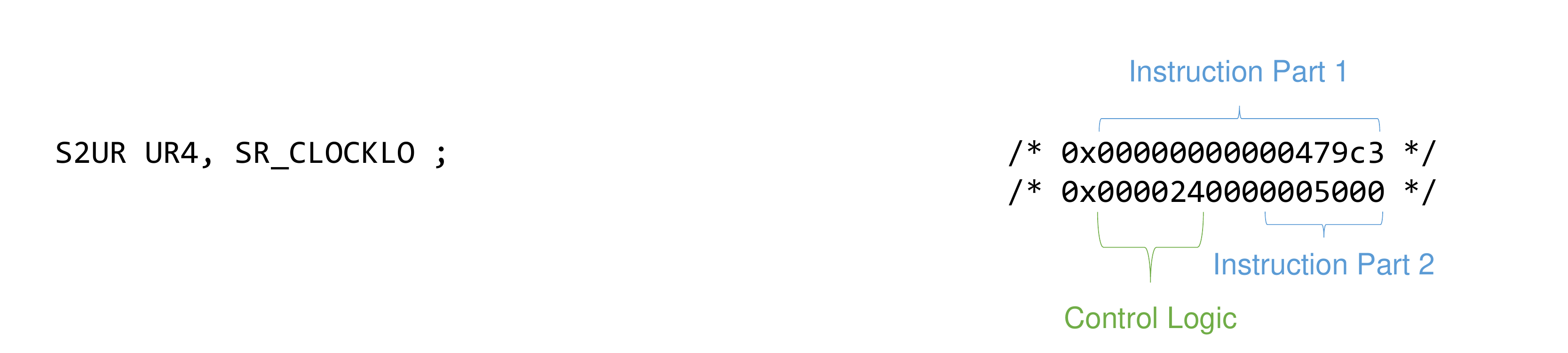} The
output shows the instruction decoded into two 64-bit words. The first
word contains pure instruction information, while the second contains
both instruction and control information.

Our experiments based on instruction disassembly and arbitrary code
execution suggest that the encoded 128 bits are used as follows:
\begin{itemize}
  \item at least 91 bits for instruction information;
  \item at least 23 bits for control information;
  \item the remaining 14 bits appeared to be unused in both Turing and
    Volta; in our experiments, they were ignored both by
    \texttt{cuobjdump} and by the hardware.
\end{itemize}

\section{Control information}

Control words appeared first with the Kepler architecture, which
substantially replaced dynamic hardware scheduling with static
software scheduling.  Control words encode instruction scheduling
decisions taken by the compiler~\cite{zhang2017} that the hardware
must enforce.  The design choice to use software scheduling in Kepler
was a departure from the previous design (Fermi): designers replaced a
complex hardware scheduler with a simpler, more efficient one that
occupied less silicon area and consumed less power.  Overall, software
scheduling enabled simpler on-chip control logic, leading to higher
compute density per area of silicon and better energy efficiency.

On Turing and Volta, 128 bits contain one instruction together with
the control information associated with only that instruction.

Pre-Volta architectures pack one control word with multiple
instruction words into a \emph{bundle}.  In each bundle, the first
word contains control information, and the remaining words (3 on
Pascal and Maxwell, 7 on Kepler) encode one instruction each. Each
control word affects how the architecture schedules the instructions
within the bundle.

The following excerpt shows a bundle of Pascal instructions decoded by
\texttt{nvdisasm}.  The bundle contains four 64-bit words.  The first
word, which has a hexadecimal dump but no corresponding disassembled
instruction, is a control word. The remaining three words are
instructions.
\begin{lstlisting}[basicstyle={\scriptsize\ttfamily} ]
                                                        /* 0x000f8800fe2007f1 */
 /*0288*/         @P5 LDG.E.CI R66, [R86+0x100];        /* 0xeed4a00010055642 */
 /*0290*/        @!P5 MOV R66, RZ;                      /* 0x5c9807800ffd0042 */
 /*0298*/         @P6 LDG.E.CI R67, [R86+0x180];        /* 0xeed4a00018065643 */
\end{lstlisting}

\noindent Control information is encoded as follows on the different
GPU generations:
\begin{itemize}
  \item on Kepler, each control word contains 6 zeroes as its most
    significant bits, 2 zeroes as its least significant bits, and 7
    sections of 8 bits each;
  \item on Pascal and Maxwell, each control word contains one zero as
    its most significant bit, and 3 sections of 21 bits each;
  \item on Turing and Volta, each control section contains 2 zeroes as
    its most significant bits, and 1 section of 21 bits. For every 128
    bits corresponding to one instruction, control information is
    preceded and followed by bits encoding the instruction itself.
\end{itemize}

Sections containing control information are organized in the same way
on Turing, Volta, Pascal and Maxwell.  Each section contains 6 fields,
organized as follows:

\begin{center}
\footnotesize
\begin{tabular}{lllllll}
  \toprule
  Width (bits)   & 4      & 6       & 3       &    3     & 1     & 4    \\
  \midrule
  Meaning        & Reuse  & Wait    & Read    & Write    & Yield & Stall \\
                 & flags  & barrier & barrier & barrier  & flag  & cycles \\
                 &        & mask    & index   & index    &       &  \\
  \bottomrule
\end{tabular}
\end{center}

\noindent Fields have the following meaning:

\textbf{Reuse flags.} Each hardware thread on Turing, Volta, Pascal
and Maxwell has a 2-way associative Content-Addressable Memory (CAM)
for each of the four conceptual source registers operand
positions. This memory is intended to allow data reuse between
instructions without accessing any register ports: this relieves
pressure on the register file, and helps reducing register bank
conflicts (we discuss register bank conflicts at length in
Section~\ref{sec:registerbanks}).  Reuse flags control this mechanism
as follows: an instruction may flag for saving into the reuse set any
combination of up to its first four arguments. Each instruction will
attempt to service register reads for its first 4 arguments from the
respective reuse slots before resorting to loading values via register
file ports. E.g., if the last two reuse-saved registers in the second
instruction source operand position were R98 and R99, either of those
registers may be used in the second position of instructions without
contributing to register bank conflicts.  The four bits in the reuse
flags map the first to fourth source operands with the least to most
significant bits, respectively.

\textbf{Wait barrier mask; Read/Write barrier index.} While most
instructions have fixed latency and can be statically scheduled by the
assembler, instructions involving memory and shared resources
typically have variable latency. Turing, Volta, Pascal and Maxwell use
\emph{dependency barriers} to track the completion of variable-latency
instructions and resolve data hazards. When a variable-latency
instruction writes to a register, the assembler associates it to one
of the 6 available barriers by setting the corresponding \emph{write
  barrier number} field. When a later instruction consumes that
register, the assembler marks the instruction as waiting on that
barrier by setting the bit corresponding to that barrier in the
\emph{wait barrier mask}. The hardware will stall the later
instruction until the results of the earlier one are available. An
instruction may wait on multiple barriers, which explains why the
\emph{wait barrier mask} is a bitmask, not an index.

\textbf{Read dependency barriers}. Read dependency barriers serve to
protect against write-after-read hazards. Unbuffered instructions that
write the contents of registers to memory need the registers to remain
unchanged during the operation. To guarantee that, the assembler
associates them to a barrier by populating the corresponding
\emph{read barrier number} field. Later instructions writing to the
same register will wait on that barrier.

\textbf{Stall cycles.} This 4-bit field indicates how long the
scheduler should wait before issuing the next instruction, ranging
from 0 to 15 cycles. On Pascal and Maxwell, if the combination of this
field and the yield flag contain a special combination of bits, the
two dispatchers in a processing block can dispatch two consecutive
instructions of a warp at the same time (dual issue). On Turing and
Volta there is only one dispatcher in a processing block, and we do
not observe dual issue in the generated code.

\textbf{Yield flag.} As its predecessors, the Turing architecture uses
a one-bit yield flag to balance the workload assigned to a processing
block.  When this bit is set, the scheduler prefers to issue the next
instruction from the current warp. When the bit is cleared, the
scheduler prefers to switch to another warp, making all register reuse
flags for the next instruction ineffective. This costs one extra cycle
to switch to another warp.

\begin{table}[b]
  \caption{This experiment reveals the same mapping between warps and
    schedulers on Turing and Volta: warps with the same index modulo 4
    are mapped to the same scheduler.  We vary the indices of two
    active warps (A and B) and measure their aggregate throughput.
    When the indices collide modulo 4 (i.e., they are mapped to the
    same scheduler) performance drops.  All values are in
    single-precision GFLOPS.}
  \label{tab:sche-flops}
  \center
  \footnotesize
  \begin{tabular}{llcccc}
    \toprule
    \multicolumn{6}{c}{T4 GPU} \\[4pt]
                                  &             & \multicolumn{4}{c}{Warp A Index} \\
                                  &             & 0     & 1      & 2    & 3           \\
    \midrule
    \multirow{4}{*}{Warp B index} &    4        & \textbf{48.9}  & 72.4   & 72.5 & 73.1 \\
                                  &    5        & 73.4  & \textbf{46.7}   & 72.5 & 73.1 \\
                                  &    6        & 73.2  & 72.8   & \textbf{47.0} & 73.2 \\
                                  &    7        & 72.9  & 72.7   & 72.7 & \textbf{46.2} \\

    \midrule
                                  &             &       &        &      &          \\
                                  &             &       &        &      &          \\
    \multicolumn{6}{c}{V100 GPU} \\[4pt]
                                  &             & \multicolumn{4}{c}{Warp A Index} \\
                                  &             & 0     & 1      & 2    & 3  \\
    \midrule
    \multirow{4}{*}{Warp B index} &    4        & \textbf{42.27} & 66.05 & 66.04 & 65.29 \\
                                  &    5        & 66.05 & \textbf{41.98} & 66.04 & 66.04 \\
                                  &    6        & 66.02 & 66.04 & \textbf{42.06} & 66.04 \\
                                  &    7        & 66.04 & 66.04 & 66.02& \textbf{42.08} \\

    \bottomrule
  \end{tabular}
\end{table}

\section{Processing Blocks and Schedulers}

The Turing streaming multiprocessor (SM) is partitioned into four
\emph{processing blocks}, each containing a dedicated warp scheduler
and dispatch unit~\cite{tu104}. Instructions from the same warp are
allocated to a specific processing block, and can only access the
processing units within that block.

We found that warps are mapped to schedulers (and processing blocks)
on Turing and Volta according to the same, simple rule:
$$ scheduler\_id = warp\_id\%4.$$
This is demonstrated with a benchmark
composed of 8 warps running on a single SM simultaneously, of which
only 2 are active with loops of FFMA instructions, while the remaining
6 are idle.

We repeat the experiments varying the warp index of each of the two
active warps (Warp A and B), while measuring each time the aggregate
arithmetic throughput achieved by the two warps. The results (see
Table~\ref{tab:sche-flops}) show that whenever the two warps have the
same index modulo 4 (e.g., 0 and 4, 1 and 5, ...), their aggregate
performance drops, which suggests that they are mapped to the same
scheduler.

These findings are consistent between Turing and Volta.

Furthermore, these results indicate that every block of your workload
must use at least 128 threads to fully utilize the processing units on
one SM of Turing and Volta.

\section{Instruction word format}

\subsection{Opcodes}

Turing and Volta use more bits to encode their instructions than in
previous architectures.

Unlike previous architectures (Pascal, Maxwell and Kepler), which
organize the opcode in the most significant bits of the instruction,
Turing and Volta architectures place the opcode in the least
significant bits of the first 64-bit word of the instruction.  Turing
opcodes vary in length from 10 to 13 bits.  For an extensive opcode
reference that compares Pascal, Volta and Turing, see the Appendix.

\subsection{Operands}
As in previous architectures, instruction operands on Turing can be
registers of different types, memory addresses (constant, shared or
global), or an immediate value.  Predication is regulated by 4 bits:
the first bit is a negation flag, and the remaining 3 bits encode a
predicate register index.

\chapter{Memory hierarchy}

\begin{figure}
  \includegraphics[width=0.96\textwidth]{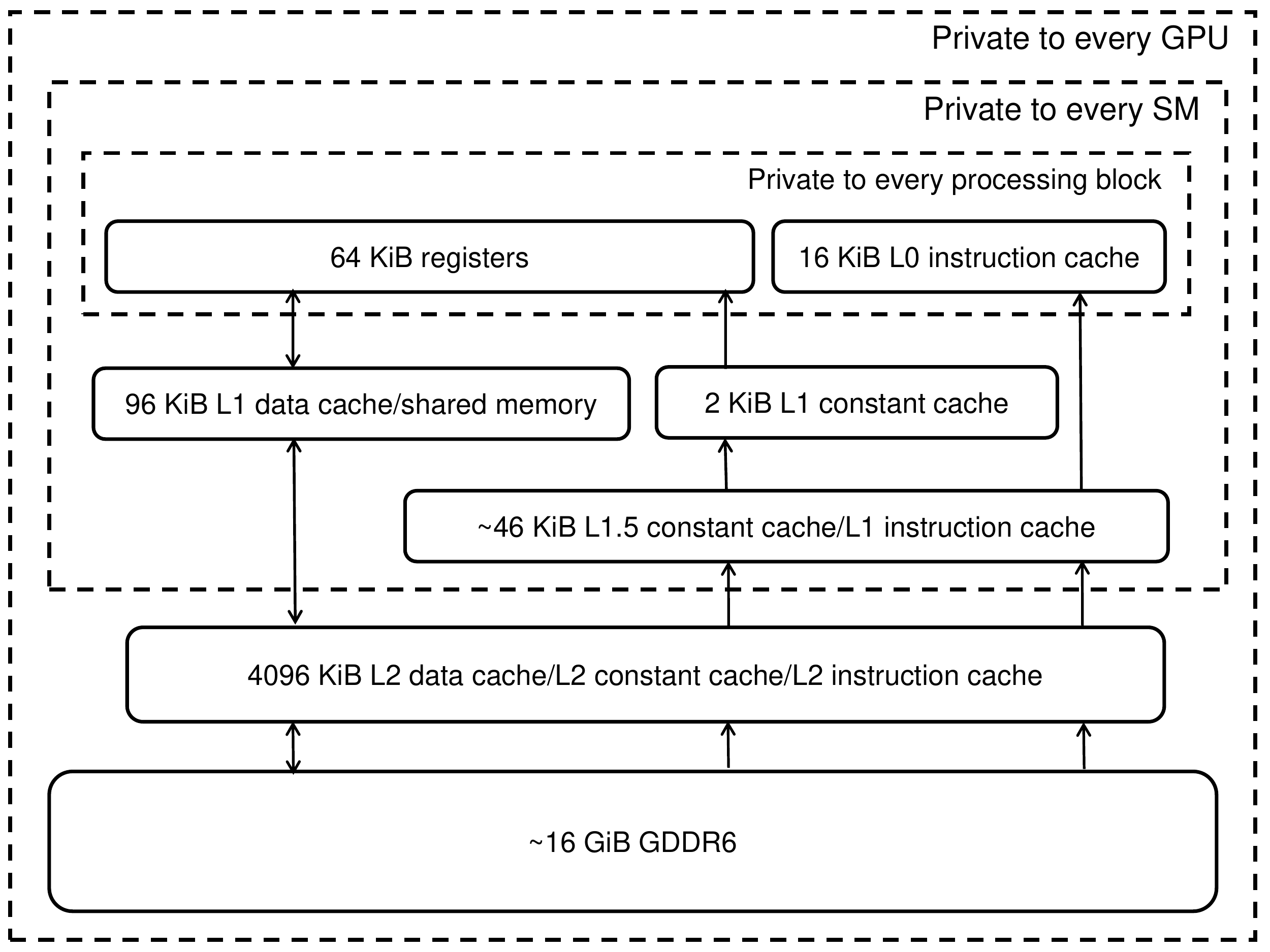}
  \caption{Memory hierarchy of the Turing T4 GPU (TU104).}
  \label{fig:memheirT4}
\end{figure}

\begin{figure}
  \includegraphics[width=0.96\textwidth]{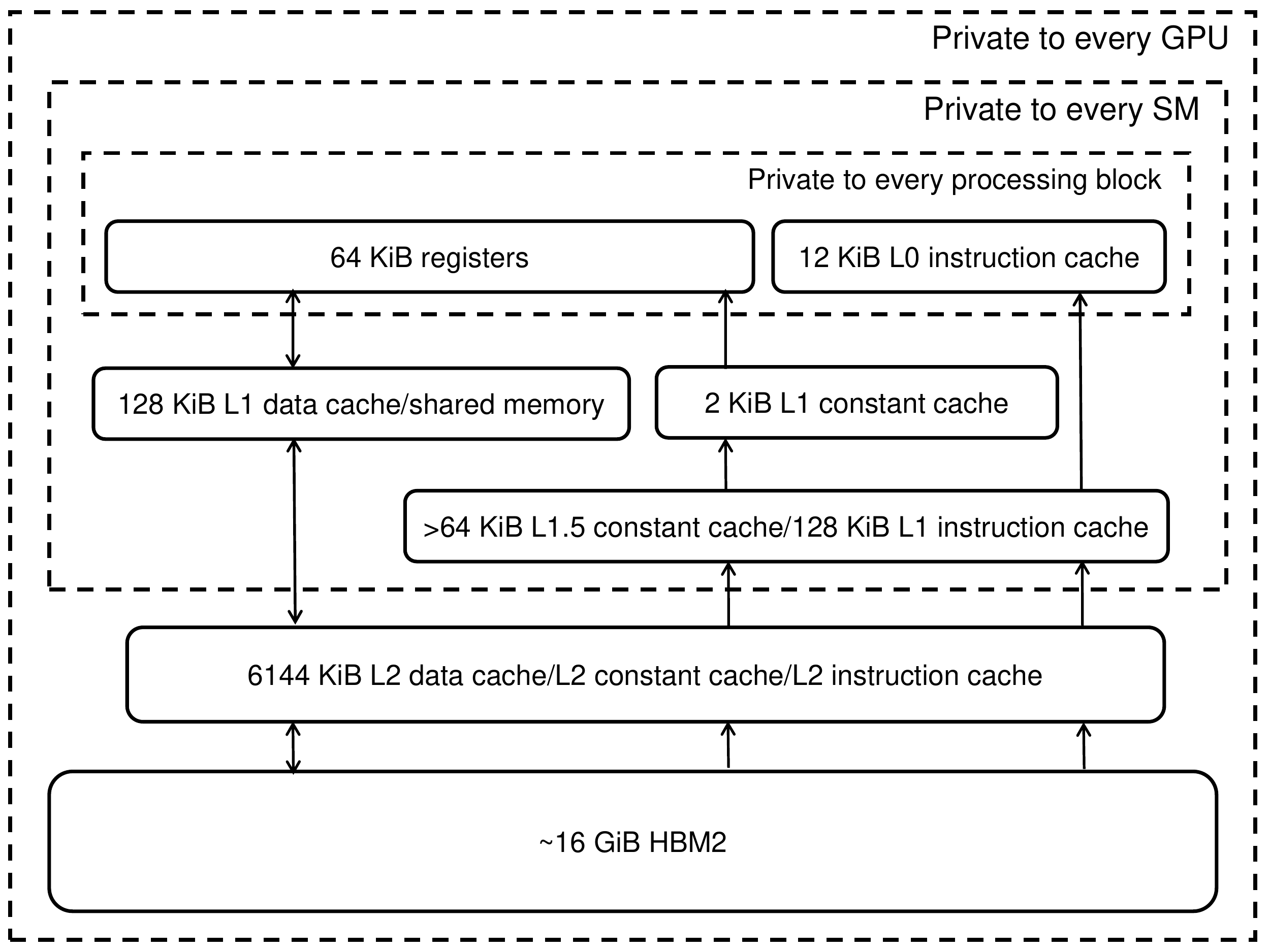}
  \caption{Memory hierarchy of the Volta V100 GPU (GV100).}
\end{figure}

\begin{figure}
  \includegraphics[width=0.96\textwidth]{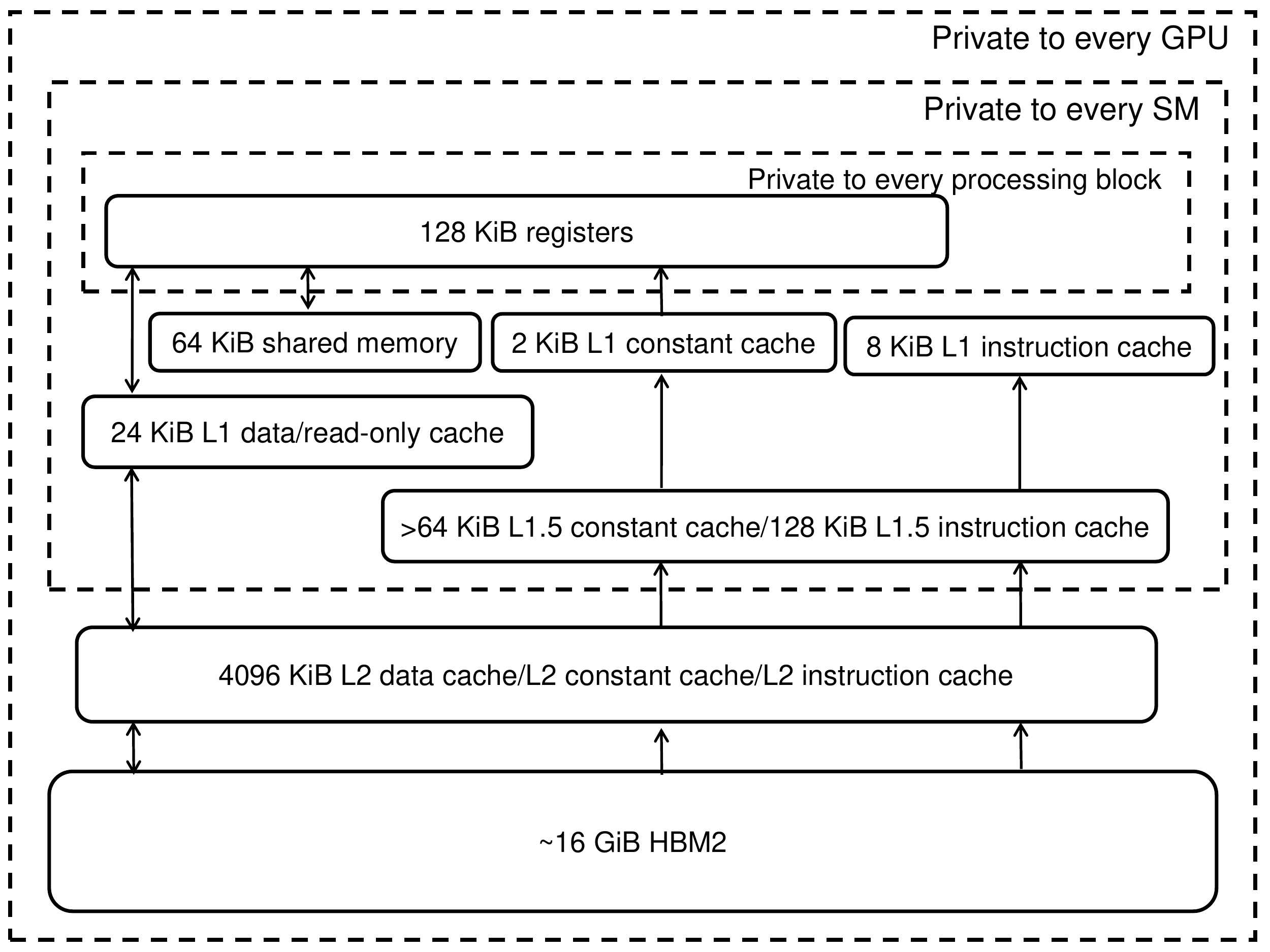}
  \caption{Memory hierarchy of the Pascal P100 GPU (GP104).}
\end{figure}

\begin{figure}
  \includegraphics[width=0.96\textwidth]{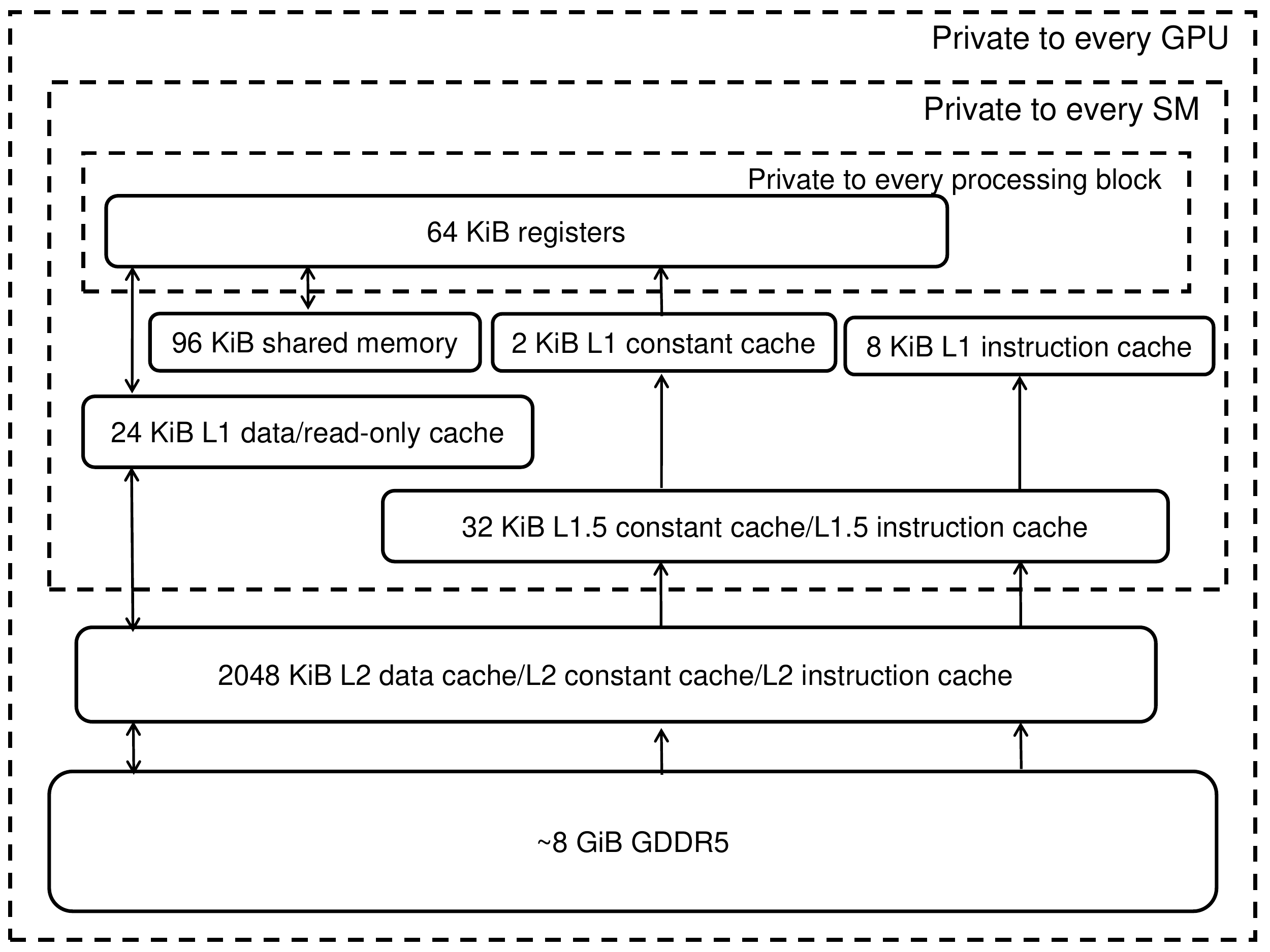}
  \caption{Memory hierarchy of the Maxwell M60 GPU (GM204).}
\end{figure}

NVidia GPU architectures tend to increase in complexity with newer
generations. Gaining a deep understanding of GPU memory hierarchy as
they evolve is necessary to write efficient code.

For designers to map their working sets optimally onto the memory
hierarchy, it is especially important to know the size of each cache
level, whether that memory is co-located with another cache
that might evict its contents, and whether each cache memory is
private to a streaming multiprocessor or shared among all.

In this chapter, we describe the structure of Turing's memory hierarchy
in detail (Figure~\ref{fig:memheirT4}). Specifically, we reveal:
\begin{itemize}
\item the geometry, properties and performance of all cache levels
  and Translation Look-aside Buffers (TLBs);
\item register file banks and their conflicts;
\item the performance of shared and global memory under load.
\end{itemize}

Table~\ref{tab:mem-hierarchy} summarizes our findings, also comparing
Turing against the Volta, Pascal, Maxwell and Kepler generations.

\begin{figure}
  \includegraphics[width=\columnwidth]{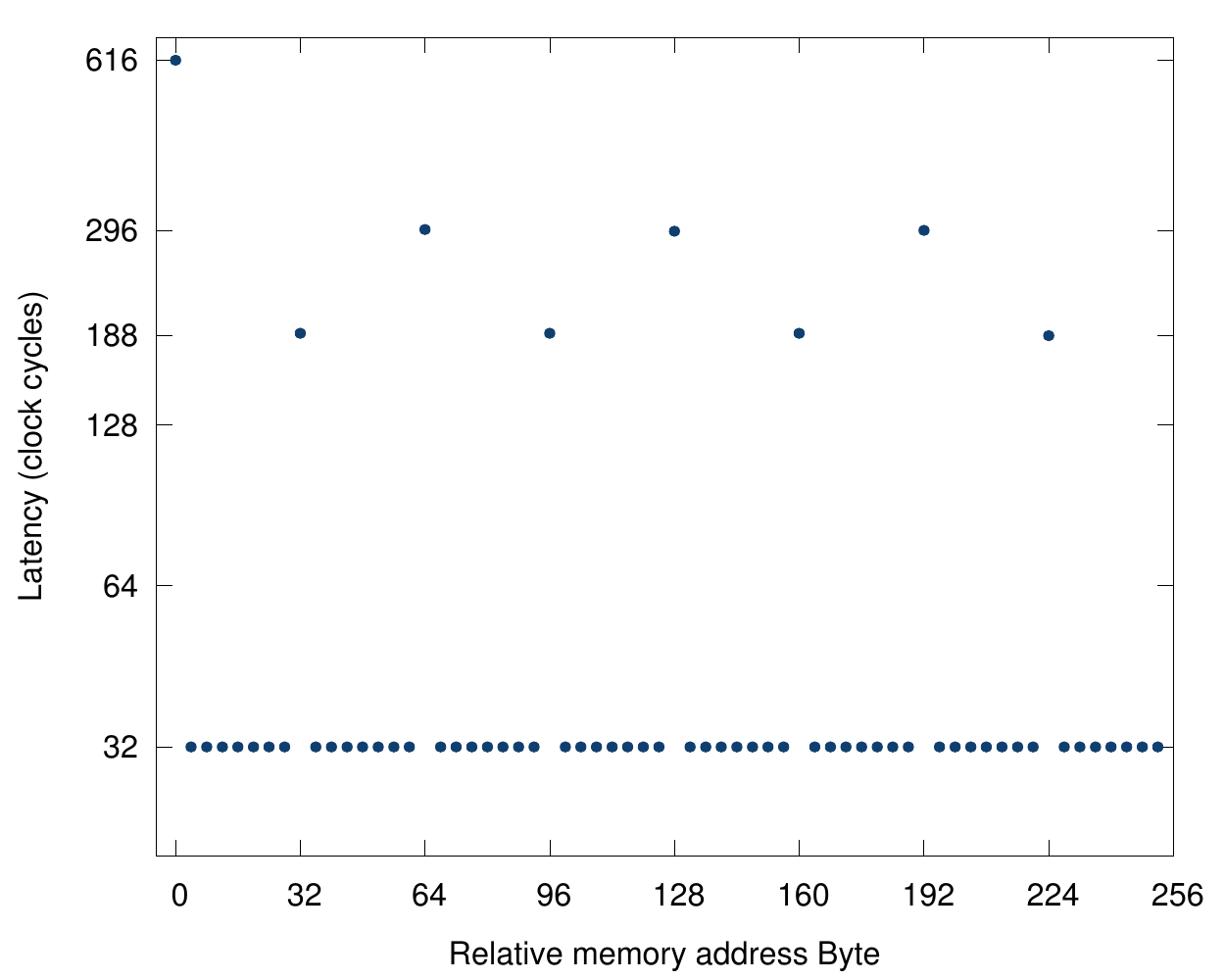}
  \caption{Global memory access latency, as per our measurements with
    the fine-grained p-chase method by Mei and Chu~\cite{mei2017}.
    The 616-cycle latency of the first access is the result of both L2
    cache miss and TLB miss. Accesses to the following data, which are
    stored in the same L1 cache line, enjoy a low L1 hit latency of 32
    cycles. Data points with a 188-cycle latency correspond to an L1
     miss and L2 hit; points with a 296-cycle latency
    correspond to an L2 miss and a TLB hit.}
  \label{fig:l1-l2-latency}
\end{figure}

The T4 GPU employs GDDR6 memory, which offers a bandwidth of 320 GB/s
(at a memory clock frequency 5,001 MHz), in conjunction with a L2
cache of 4,096 KiB~\cite{tu104}.  Data loaded from global memory is
implicitly cached in L1 and L2.

\begin{table}[t]
  \caption{Geometry, properties and latency of the memory hierarchy
    across GPU architectures. For consistency, all performance data in
    this table were measured on PCI-E cards.}
  \label{tab:mem-hierarchy}
  \setlength{\tabcolsep}{2pt}
  \center
  \footnotesize
  \makebox[\textwidth]{%
    \centering
    \begin{tabular}{lllrrrrrr}
  \             & Architecture generation         &      & Turing                    & Volta         & Pascal        & Pascal        & Maxwell    & Kepler    \\
  \             & GPU Board                       &      & T4                        & V100          & P100          & P4            & M60        & K80       \\
  \             & GPU Chip                        &      & TU104                     & GV100         & GP100         & GP104         & GM204      & GK210     \\
  \midrule
  \             & Processors per chip ($P$)       &      & 40                        & 80            &  56           & 40            & 16            & 13       \\
  \             & Max graphics clock  ($f_g$)     & MHz  & 1,590                     & 1,380         &  1,328        & 1,531         & 1,177         & 875      \\
  \             & Threads per Multiprocessor      &      & 1,024                     & 2,048         &  2,048        & 2,048         & 2,048         &2,048     \\
  \midrule
  Registers     &  Number of banks                &      & 2                         & 2             & 4             & 4             & 4             &  4            \\
                &  Bank width                     & bits & 64                        & 64            & 32            & 32            & 32            & 32            \\
  \midrule
  L1 data       &  Size                           & KiB    & 32 or 64                  & 32...128      & 24            & 24            & 24            & 16...48        \\
  \             &  Line size                      & B      & 32                        & 32            & 32            & 32            & 32            & 128          \\
  \             &  Hit latency                    & cycles & 32                        & 28            & 82            & 82            & 82            & 35           \\
  \             &  Load granularity               & B      & 32                        & 32            & 32            & 32            & 32            & 128          \\
  \             &  Update granularity             & B      & 128                       & 128           & 128           & 128           & 128           & 128          \\
  \midrule
  L2 data       &  Size                           & KiB    & 4,096                     & 6,144         & 4,096         & 2,048         & 2,048         & 1,536     \\
  \             &  Line size                      & B      & 64                        & 64            & 32            & 32            & 32           & 32         \\
  \             &  Hit latency                    & cycles & $\sim$188                 & $\sim$193     & $\sim$234     & $\sim$216     & $\sim$207     & $\sim$200   \\
  \midrule
  L1 const      &  Broadcast latency              & cycles & $\sim$26                  & $\sim$27      & $\sim$24      & $\sim$25      & $\sim$25      & $\sim$30   \\
                &  Cache size                     & KiB    & 2                         & 2             & 2             & 2             & 2             & 2      \\
                &  Line size                      & B      & 64                        & 64            & 64            & 64            & 64            & 64        \\
                &  Number of sets                 &        & 8                         & 8             & 8             & 8             & 8             & 8         \\
                &  Associativity                  &        & 4                         & 4             & 4             & 4             & 4             & 4         \\
  L1.5 const    &  Broadcast latency              & cycles & 92                        & $\sim$89      & $\sim$96      & $\sim$87      & $\sim$81      & $\sim$92   \\
                &  Cache size                     & KiB    & $\sim$46                  & $>=$64        & $>=$64        & 32            & 32            & 32         \\
                &  Line size                      & B      & 256                       & 256           & 256           & 256           & 256           & 256        \\
  L2 constant   &  Broadcast latency              & cycles & $\sim$215                 & $\sim$245     & $\sim$236     & $\sim$225     & $\sim$221     & $\sim$220  \\
  \midrule
  L0 instruction   &  Cache size                  & KiB  & $\sim$16                  & $\sim$12      & -             & -             & -             & -          \\
  L1 instruction   &  Cache size                  & KiB  & $\sim$46                  & 128           & 8             & 8             & 8             & 8          \\
  L1.5 instruction &  Cache size                  & KiB  & -                         & -             & 128           & 32            & 32            & 32         \\
  L2 instruction   &  Cache size                  & KiB  & 4,096                     & 6,144         & 4,096         & 2,048         & 2,048         & 1,536      \\
  \midrule
  L1 TLB        &  Coverage                       & MiB  & 32                        & 32            & $\sim$32      & $\sim$32      & $\sim$2       & $\sim$2    \\
                &  Page entry                     & KiB   & 2,048                    & 2,048         & 2,048         & 2,048         & 128           & 128        \\
  L2 TLB        &  Coverage                       & MiB  & $\sim$8,192               &$\sim$8,192    &$\sim$2,048    &$\sim$2,048    &$\sim$128      &$\sim$128   \\
                &  Page entry                     & MiB  & 32                        & 32            & 32            & 32            & 2             & 2          \\
  L3 TLB        &  Coverage                       & GiB  & -                         & -             & -             & -             & $\sim$2       &$\sim$2     \\
                &  Page entry                     & MiB  & -                         & -             & -             & -             & 2             & 2          \\
  \midrule
  Shared        &  Size per SM                    & KiB  & 32 or 64                  & 0...96        & 64            & 64            & 96            & 48      \\
  \             &  Size per chip                  & KiB  & 1,280 or 2,560            & 0...7,689     & 3,584          & 1,280       & 1,536         & 624         \\
  \             &  Banks per processor ($B_s$)    &      & 32                        &            32  &            32  &         32  &           32  &         32  \\
  \             &  Bank width ($w_s$)             & B     & 4                        &            4  &           4   &        4   &          4   &        8   \\
  \             &  LSU count per SM  ($n_{LSU}$)   &      & 16                        &            32  &           16  &        16   &          32   &        32  \\
  \             &  No-conflict latency            & cycles & 19                        &             19 &             24 &           23 &             23 &          26  \\
  \             &  Theoretical bandwidth          & GiB/s & 4,070                     & 13,800         & 9,519          & 3,919        & 2,410         & 2,912        \\
  \             &  Actual bandwidth               & GiB/s & 3,662                     & 12,080         & 7,763          & 3,555        & 2,122         & 2,540        \\
  \             &  Actual/Theoretical ratio       & \%    & 90.9\%                    & 87.5\%         & 81.6\%         &   90.7\%     & 88.0\%        &    87.2\%    \\
  \midrule
  Global        &  Memory bus                     &       & GDDR6                     & HBM2           &  HBM2          & GDDR5        & GDDR5          & GDDR5       \\
  \             &  Size                           & MiB   & 15,079                    & 16,152         & 16,276         & 8,115        & 8,155          & 12,237      \\
  \             &  Max clock rate ($f_m$)         & MHz   & 5,001                     & 877            &    715         & 3,003        & 2,505          &  2,505      \\
  \             &  Theoretical bandwidth          & GiB/s & 320                       & 900            &   732          & 192          & 160            &  240        \\
  \             &  Actual bandwidth               & GiB/s & 220                       & 750            &   510          &  162         & 127            &   191        \\
  \             &  Actual/Theoretical ratio       & \%    & 68.8\%                    & 83.3\%         &      69.6\%    &       84.4\% &      79.3\%    &    77.5\%    \\
  \bottomrule
  \end{tabular}
  }
\end{table}

\section{L1 data cache}

Turing adopts the same combined L1 data cache / shared memory design
as Volta. This design reduces the cache hit latency and improves the
bandwidth with respect to the Pascal architecture.

As the geometry of the L1 data cache is concerned, our findings agree
with what reported in the Turing and Volta architecture
whitepapers~\cite{gv100,tu104}. Specifically, the T4 offers twice as
much L1 data capacity, and twice as high a bandwidth as the P4 GPU.

As performance is concerned, our experiments show that on a T4 GPU,
the L1 data cache offers approximately 3.7$\times$ more bandwidth than
its P4 predecessor.

\subsection{Latency and bandwidth}

The L1 data cache hit latency we measured on the T4 GPU is
32 cycles, compared to 82 cycles on the P4 (see
Figure~\ref{fig:l1-l2-latency}).

Before Turing and Volta, Kepler was the most recent architecture to
combine its L1 cache and its shared memory. Kepler's L1 cache read hit
latency is 35 clock cycles. Turing exhibits a better L1 latency than
Kepler in clock cycles, despite the T4 being clocked almost twice as
high as the K80 (1,590 vs. 875 MHz).

We use the following benchmark to measure the L1 data
cache load bandwidth. The benchmark scans an array with 32-bit
elements; every warp accesses all the elements in the array:
\begin{lstlisting}[basicstyle={\scriptsize\ttfamily}]
__global__ void l1_bw(  uint32_t *startClk, uint32_t *stopClk,
                        float *dsink, uint32_t *posArray )
{
    // Thread index
    uint32_t tid = threadIdx.x;

    // Side-effect variable, intended to avoid compiler elimination of this code
    float sink = 0;

    // Warm up the L1 cache by populating it
    for (uint32_t i = tid; i<L1_SIZE; i+=THREADS_NUM) {
      float * ptr = posArray+i;
      asm volatile ("{\t\n"
        ".reg .f32 data;\n\t"
        "ld.global.ca.f32 data, [%1];\n\t"
        "add.f32 %0, data, %0;\n\t"
        "}" : "+f"(sink) : "l"(ptr) : "memory"
      );
    }

    // Synchronize all threads
    asm volatile ("bar.sync 0;");

    // Start timing
    uint32_t start = 0;
    asm volatile ("mov.u32 %0, %%clock;" : "=r"(start) :: "memory");

    // Load data from L1 cache, accumulate
    for (uint32_t i = 0; i<L1_SIZE; i+=THREADS_NUM) {
      float * ptr = posArray+i;
      // every warp loads all data in l1 cache
      for (uint32_t j = 0; j<THREADS_NUM; j+=WARP_SIZE) {
        uint32_t offset = (tid+j)%THREADS_NUM;
        asm volatile ("{\t\n"
          ".reg .f32 data;\n\t"
          "ld.global.ca.f32 data, [%1];\n\t"
          "add.f64 %0, data, %0;\n\t"
          "}" : "+f"(sink) : "l"(ptr+offset) : "memory"
        );
      }
    }

    // Synchronize all threads
    asm volatile ("bar.sync 0;");

    // Stop timing
    uint32_t stop = 0;
    asm volatile ("mov.u32 %0, %%clock;" : "=r"(stop) :: "memory");

    // Write time and data back to memory
    startClk[tid]  = start;
    stopClk[tid]   = stop;
    dsink[tid]     = sink;
}
\end{lstlisting}

\begin{table}[b]
  \caption{L1 cache load throughput per SM.}
  \label{tab:l1-bw}
  \center
  \footnotesize
  \begin{tabular}{lrrrrrrl}
  \toprule
                           & T4      & V100   & P100 & P4   & M60  & K80 \\
  \midrule
  Theoretical upper bound  & 64.0    & 128.0  & 64.0  & 64.0  & 128.0 & 128.0 &  bytes/cycle \\
  Measured throughput      & 58.8    & 108.3  & 31.3  & 15.7  & 15.7  & 68.6  &  bytes/cycle \\
  \bottomrule
  \end{tabular}
\end{table}

We report L1 data bandwidths we measured across GPU devices in
Table~\ref{tab:l1-bw}, together with their theoretical upper bounds.

The actual bandwidth we measure on the T4 GPU is 58.83 bytes/cycle/SM,
i.e., 3.7$\times$ higher than that of the P4 GPU, i.e., 15.7
bytes/cycle/SM. This bandwidth comparison expressed in cycle counts is
meaningful, because the T4 and P4 cards run at very similar graphics
clock frequencies $f_g$.

We calculate the theoretical throughput by multiplying the LSU count
per SM by the number of bytes that each LSU can load per cycle per
instruction.

Historically, architectures that employ an L1 cache combined with shared
memory (Turing, Volta and Kepler) exhibit a higher L1 bandwidth than architectures where
the L1 cached and the shared memory are separate (Pascal and Maxwell).

\subsection{Geometry}

According to the Turing whitepaper~\cite{tu104}, load/store operations
can use a L1 data cache of 32 KiB or 64 KiB in size.

Our experiments based on Mei and Chu's fine-grained pointer-chase
technique~\cite{mei2017} were unable to detect the whole configured
size, and fell 7 KiB short of the nominal L1 data cache size, on both
Volta and Turing architectures (see Table~\ref{tab:l1-size}).

In our experimental setup, the shared memory is configured to a size
of 64 KiB.  We then employed a benchmark that scans a variable length
array \textbf{A} twice. As long as the size of \textbf{A}
exceeds 25 KiB, we detected cache misses.

At this time we are unable to explain this 7-KiB discrepancy. We
conjecture it is the result of a newly applied replacement policy that
we discuss below. We confirm that it is not associated to the ECC
feature (error correction).

\begin{table}
  \caption{Detectable L1 data cache size with the pointer-chase
    benchmark on the T4 GPU.}\label{tab:l1-size}
  \center
  \footnotesize
  \begin{tabular}{lrr}
  \toprule
  Configured size of shared memory (KiB) & 32   & 64  \\
  \midrule
  Expected size of L1 data cache (KiB)   & 64   & 32   \\
  Detected size of L1 data cache (KiB)   & 57   & 25   \\
  \bottomrule
  \end{tabular}
\end{table}

Table~\ref{tab:mem-hierarchy} describes the remainder of L1 data cache
geometry as we discover it. The line size, load and update granularity
of Turing's L1 data cache are the same as on the Volta, Pascal and
Maxwell GPUs.

In our previous report for Volta~\cite{zhe2018}, we discovered an
improved L1 cache replacement policy on Volta with respect to its
predecessors.  Turing also features a L1 cache replacement policy that
aims at preserving large arrays from eviction caused by sparse memory
accesses.

We employed a benchmark that scans a variable length array twice, and
recorded positions and latency data when L1 cache miss happens. We
found that when the L1 data cache saturates, Turing randomly evicts 4
consecutive cache lines (128 B). We observed that once a block of
cache lines are evicted, the second scan will cause more cache lines
from the same set to be evicted.

\section{Unified L2 cache}

Turing employs an L2 cache that is unified for data, instructions and
constant memory, as the previous GPU generations do.  The L2 cache on
a T4 GPU is a 16-way, set-associative cache having a size of 4,096
KiB, a cache line of 64 B, and an average latency of 188 clock cycles
(Figure~\ref{fig:l1-l2-latency}).

\begin{table}
  \caption{L2 data cache load throughput.}
  \label{tab:l2-bw}
  \center
  \footnotesize
  \begin{tabular}{lcccccc}
    \toprule
    \                  & Turing & Volta & Pascal & Pascal & Maxwell & Kepler \\
    \                  &  T4    & V100  & P100   &  P4    &  M60    &  K80 \\
    \midrule
    Throughput (GB/s)  & 1,270  & 2,155 & 1,624  &   979  &     446 & 339    \\
    \bottomrule
  \end{tabular}
\end{table}

We use the following benchmark to measure L2 load bandwidth, on all
the GPUs considered:
\begin{lstlisting}[basicstyle={\scriptsize\ttfamily}]
__global__ void l2_bw(float *dsink, uint32_t *posArray)
{
    // block and thread index
    UINT tid = threadIdx.x;
    UINT bid = blockIdx.x;

    // accumulator; side effect to prevent code elimination
    float sink = 0;

    // load data from l2 cache and accumulate
    for (UINT i = 0; i<L2_SIZE; i+=THREADS_NUM) {
      DTYPE* ptr = posArray+i;
      // every warp loads all data in l2 cache
      for ( UINT j=0; j < THREADS; j+=32 ){
        UINT offset = (tid+j)%THREADS;
        asm volatile ("{\t\n"
          ".reg .f32 data;\n\t"
          "ld.global.cg.f32 data, [%1];\n\t"
          "add.f32 %0, data, %0;\n\t"
          "}" : "+f"(sink) : "l"(ptr+offset) : "memory"
        );
      }
    }

    // side effect: store the result
    dsink[tid] = sink;
}
\end{lstlisting}

Note that we warm up the L2 cache before launching this kernel (code
not shown for brevity).  The benchmark contains a simple
floating-point accumulation into variable \texttt{sink}, which is
later written to global memory; this accumulation intentionally
creates a side effect intended to prevent the compiler from
eliminating the entire benchmark code.  The marginal cost of this
accumulation is negligible with respect to the data access latency.

The bandwidth we measure on the T4 device (see results in
Table~\ref{tab:l2-bw}) is 30\% higher than the P4's, and 59\% of the
one measured on the larger V100 GPU.

\begin{landscape}
\begin{figure}
  \begin{center}
    \includegraphics[width=1.6\textwidth]{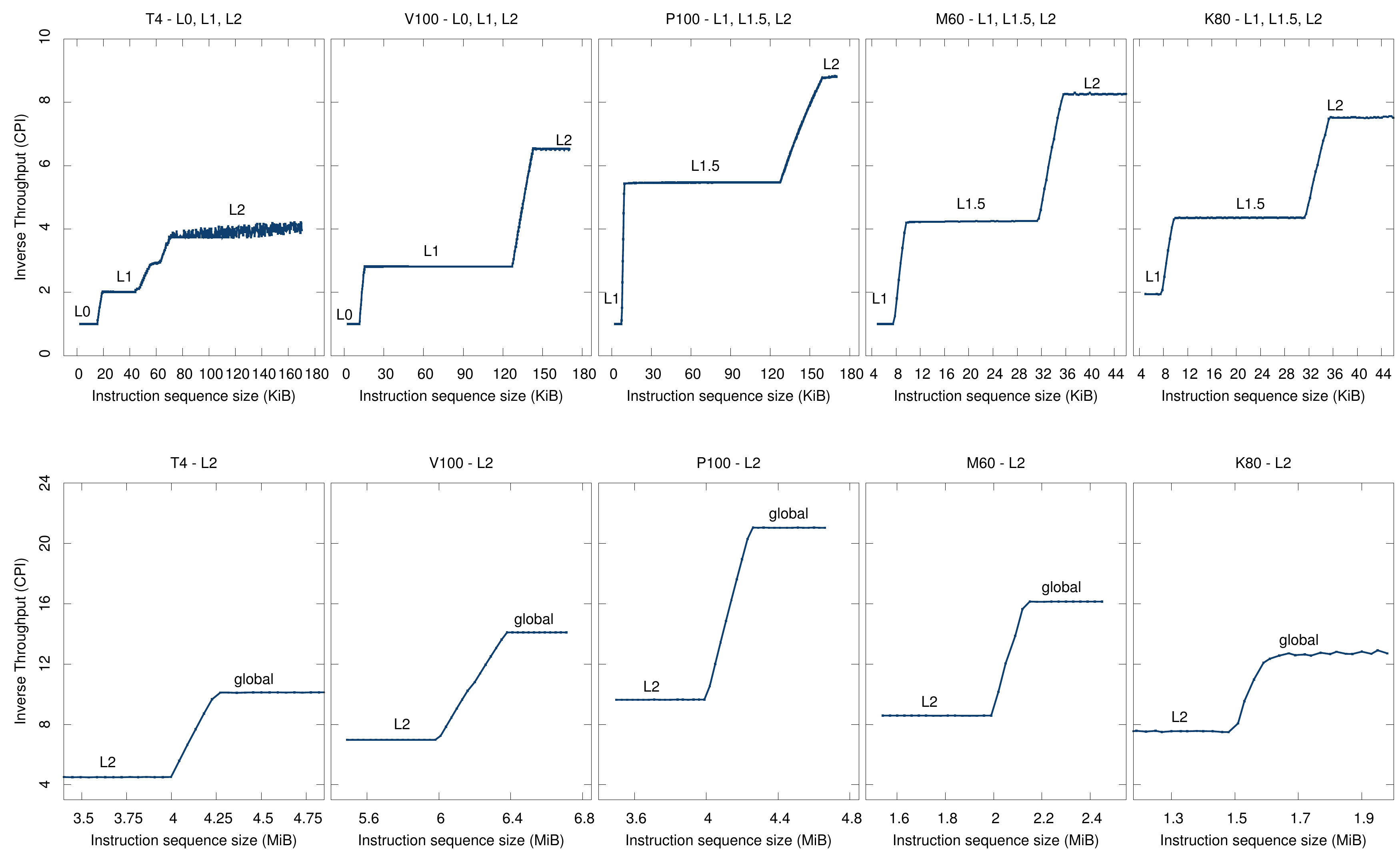}
  \end{center}
  \caption{We detect the size of instruction cache level with a
    benchmark based on sequences of identical instructions of
    increasing length. We then chart the average inverse throughput:
    each plateau reveals the size of a cache level.  \textbf{Top
      charts:} boundaries of the first two levels in the hierarchy.
    \textbf{Bottom charts:} limits of the last level, misses to global
    memory.}
  \label{fig:inst-size}
\end{figure}
\end{landscape}

\section{Instruction cache hierarchy}

In this section, we map experimentally the size and the organization
of the instruction cache hierarchy. In practice, that consists in (1)
detecting the size of each cache level and (2) determining how cache
levels are distributed within the architectural blocks (scheduler,
SM, entire chip) of the GPU.

\subsection{Taxonomy}
All GPU architectures we considered, including
Turing, feature three levels of instruction caches.  To avoid
confusion, note that on Turing and Volta the three levels are named
differently (L0, L1, L2) than on previous architectures (L1, L1.5,
L2). We adopt this established taxonomy for consistency with the
NVidia's whitepapers~\cite{tu104,gv100} and with prior literature.
Pay attention to expressions like ``the second level of instruction
caches'': this expression refers to L1 on Turing and Volta, but to L1.5
on Pascal, Maxwell and Kepler.

\subsection{Size}
To detect the size of each cache level, we study how the average
inverse throughput (i.e., average clocks per instruction, or CPI)
achieved by a long sequence of instructions changes as a function of
sequence length. As we increase the length of a sequence, we expect to
see a constant CPI value until the sequence exceeds the cache
size. Indeed, experimental results show plateaus and ramps
(Figure~\ref{fig:inst-size}) which correspond to cache level sizes and
transitions from one level to the following. In the figure, the bottom
charts focus on the three instruction cache levels, whereas the bottom
charts focus on the transition between the last cache level and global
device memory.

We report all findings in Table~\ref{tab:mem-hierarchy}.  Turing
enjoys better inverse throughput than its predecessors when accessing
the second and third instruction cache levels.

{\noindent\small\textbf{Experimental setup.}  Our benchmark measures the
  average CPI seen by a sequence of instructions of given length that
  exert no pressure on the data cache hierarchy. We iterate
  measurements for sequence sizes starting from the cache line size up
  to the larger plausible size of L3.  The benchmark executes each
  sequence twice, but only times the second execution, so that we only
  measure \emph{capacity misses} and \emph{conflict misses}, but not
  \emph{cold misses}.
  \begin{itemize}
    \item On Pascal, Maxwell and Kepler, we employ the same technique
      as in our previous report~\cite{zhe2018} for the sake of
      consistency, i.e., long sequences of FFMA instructions, whose
      register operands are chosen so that each instruction
      experiences no register dependence with its neighbors.
    \item On Volta and Turing, we switched to a simpler method that
      uses NOP sequences rather than FFMA. This choice circumvents
      NVCC's undesired generation of 2-cycle stalls between consequent
      FFMA instructions on these two GPUs.
  \end{itemize}
}

\begin{figure}
  \includegraphics[width=\columnwidth]{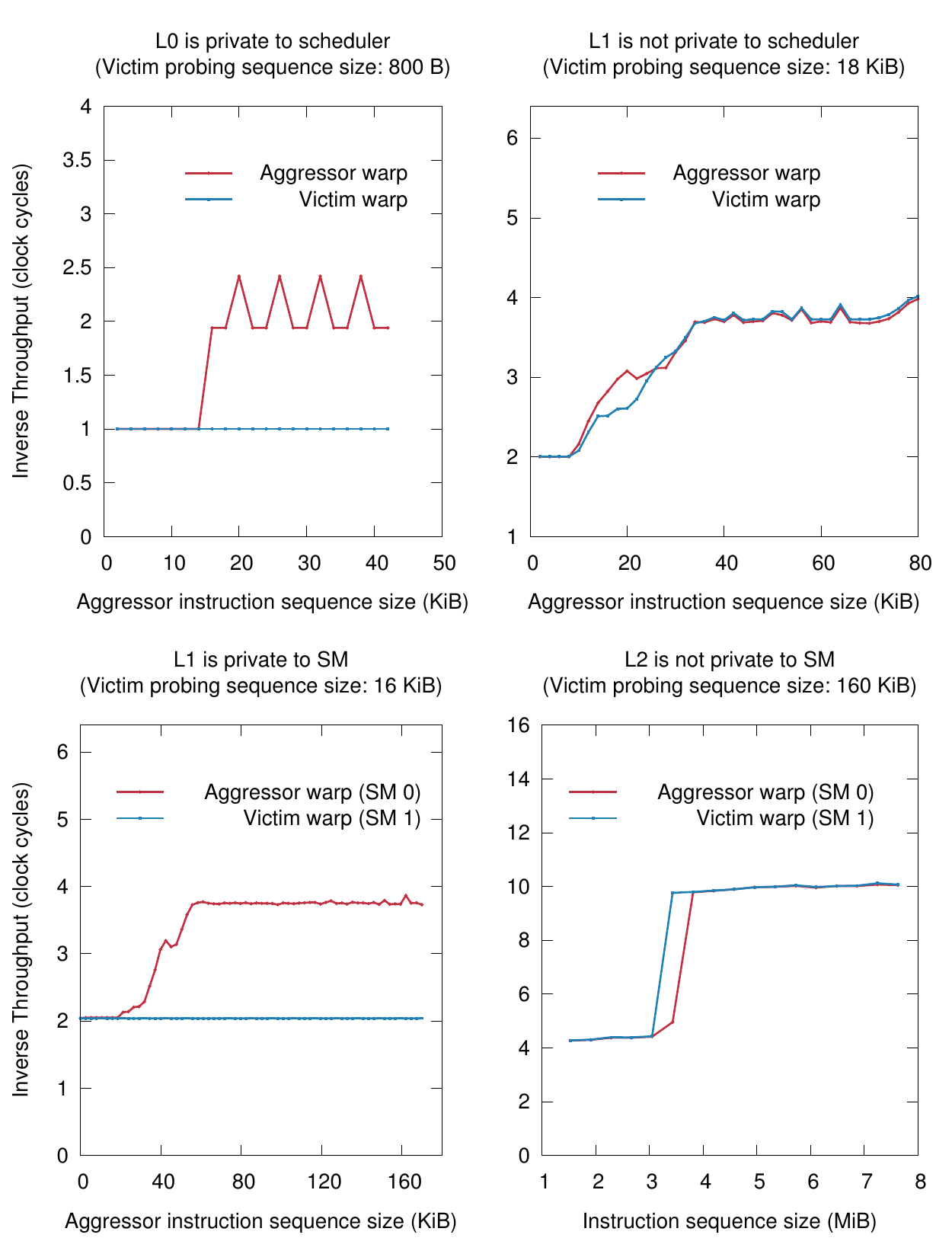}
  \caption{Aggressor-victim experiments designed to detect what
    architectural block (a scheduler, an SM, the entire chip) owns
    each level of the instruction cache, by observing how an
    aggressor warps that causes instruction cache pressure affects a
    victim warp's performance.  \textbf{Top left:} each L0 instruction
    cache is private to a scheduler.  \textbf{Top right:} an L1
    instruction cache is not private to a scheduler.  \textbf{Bottom
      left:} each L1 instruction cache is private to one SM.
    \textbf{Bottom right:} the L2 instruction cache is common among
    all SMs.}
  \label{fig:inst-private}
\end{figure}

%
%
%
%


\subsection{Organization}

Across the different GPU architectures, levels in the instruction
memory hierarchy are organized as follows:
\begin{itemize}
\item on Turing and Volta, each L0 instruction cache is private to one
  scheduler/processing block;
\item on all GPUs considered, each L1 instruction cache is private to
  an SM;
\item on Pascal, Maxwell and Kepler each L1.5 instruction cache is
  private to one SM; the L1.5 instruction cache does not exist on
  Turing and Volta;
\item on all GPUs considered, the L2 cache is unified (i.e., it caches
  instructions and data) and it is shared across all SMs.
\end{itemize}

On architectures older than Turing, we provided experimental support
for these claims in our previous report~\cite{zhe2018}.  For claims
about Turing, we collected evidence using experiments designed as
follows.

Our experiments measure the interaction between an \emph{aggressor
  warp} and a \emph{victim warp}. Both warps loop through sequences of
NOP instructions of chosen length:
\begin{itemize}
\item the \textbf{victim} warp only runs a fixed-length NOP sequence,
  typically designed to fit within a certain instruction cache level;
  we call it the \emph{victim probing sequence};
\item the \textbf{aggressor} warp runs, in addition to the same
  probing sequence as the victim, and \emph{before} it, a
  variable-length NOP sequence, designed to thrash a given cache
  level, potentially evicting instruction cache entries.
\end{itemize}
We monitor whether the evictions caused by the aggressor warp only
affect its own performance, or they affect the victim as well: if the
victim is unaffected, then the smallest cache level that fits the
fixed-length victim probing sequence is private to the architectural
block where the two warps are running (i.e., GPU processing block or
SM); else, the cache level is shared between the two warps and located
outside the block considered. In our experiments, both warps monitor
their performance by measuring their inverse throughput (CPI).

Results show that each L0 instruction cache is private to a processing
block, that each L1 instruction cache is private to an SM, and that
the L2 cache is shared among all SMs (Figure~\ref{fig:inst-private}).

To examine the relation between levels L0, L1 and \textbf{schedulers}
(or GPU processing blocks), we use experiments where the aggressor and
victim warps run on the same SM, but different processing blocks. We
use increasingly longer sequences in the aggressor warp.  To exclude
compulsory misses from the measurements, we let the aggressor and then
the victim warm up the caches by running each their respective
sequence once.

We observe that:
\begin{itemize}
\item as the aggressor sequence grows while remaining below L0
  capacity, only the aggressor experiences a slowdown (top left chart
  in Fig.~\ref{fig:inst-private}), whereas the victim is
  unaffected. This indicates that the two warps access distinct L0
  caches, private to each processing block;
\item as the instruction sequence grows above L0 capacity (top right
  chart) and into L1, both warps slow down similarly, which indicates
  that the two warps share L1.
\end{itemize}

Next, we examine the relation between levels L1 and L2, and SMs, with
similarly constructed experiments. This time, the two warp run on
separate SMs (SM0 and SM1).

We observe that:
\begin{itemize}
\item as the aggressor sequence exceeds L0 but remains within L1
  capacity, only the aggressor warp experiences a slow-down
  corresponding to L1 hit rates (bottom left); the victim, still
  running a sequence fitting L0 (16 KiB), is unaffected. This
  indicates that different SMs have distinct L1 caches;
\item as the aggressor sequence exceeds L2 capacity (bottom right
  chart), both victim and aggressor experience slowdowns; This
  indicates that different SMs access the same L2 cache.
\end{itemize}

\section{Constant memory hierarchy}

\begin{figure}
  \center
  \includegraphics[width=\columnwidth]{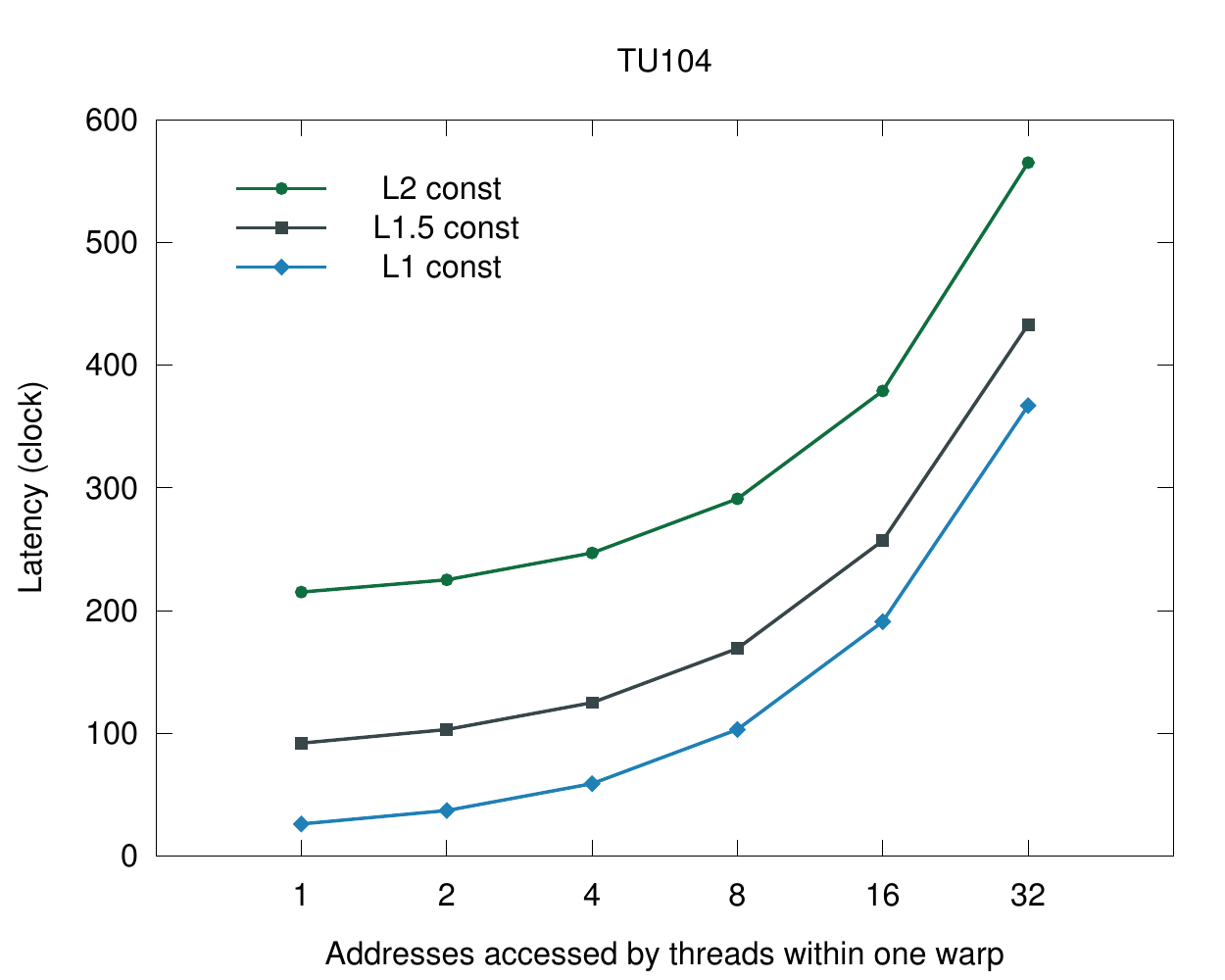}
  \caption{Latency of concurrent loads from constant memory within a warp
           depends on where the data is found in the cache hierarchy (L1,
           L1.5, or L2) and on the count of distinct locations referenced.
           The hardware broadcasts accesses to the same location.}
  \label{fig:const_latency}
\end{figure}

The constant memory is a cached window of global memory, reserved for
data declared with the \texttt{\_\_constant\_\_} keyword, plus kernel
invocation parameters and immediate constants. We find that Turing has
three levels of constant cache memory, which have the geometry and
properties described in Table~\ref{tab:mem-hierarchy} and latency as
in Figure~\ref{fig:const_latency}.

The constant memory hierarchy used in Turing did not change
significantly from previous generations.  Across all the GPU
generations we considered, the following properties hold true:
\begin{itemize}
\item the L1 constant cache uses a non-LRU replacement policy;
\item each SM possesses two private levels of constant caches, which
  we denote as L1 and L1.5 constant cache (accesses to either of each
  level within an SM do not affect the same cache levels on other SMs);
\item the L2 cache is the third level of constant cache. It is shared
  among all SMs and is unified for instruction and data.
\end{itemize}

\begin{figure}
  \includegraphics[width=\columnwidth]{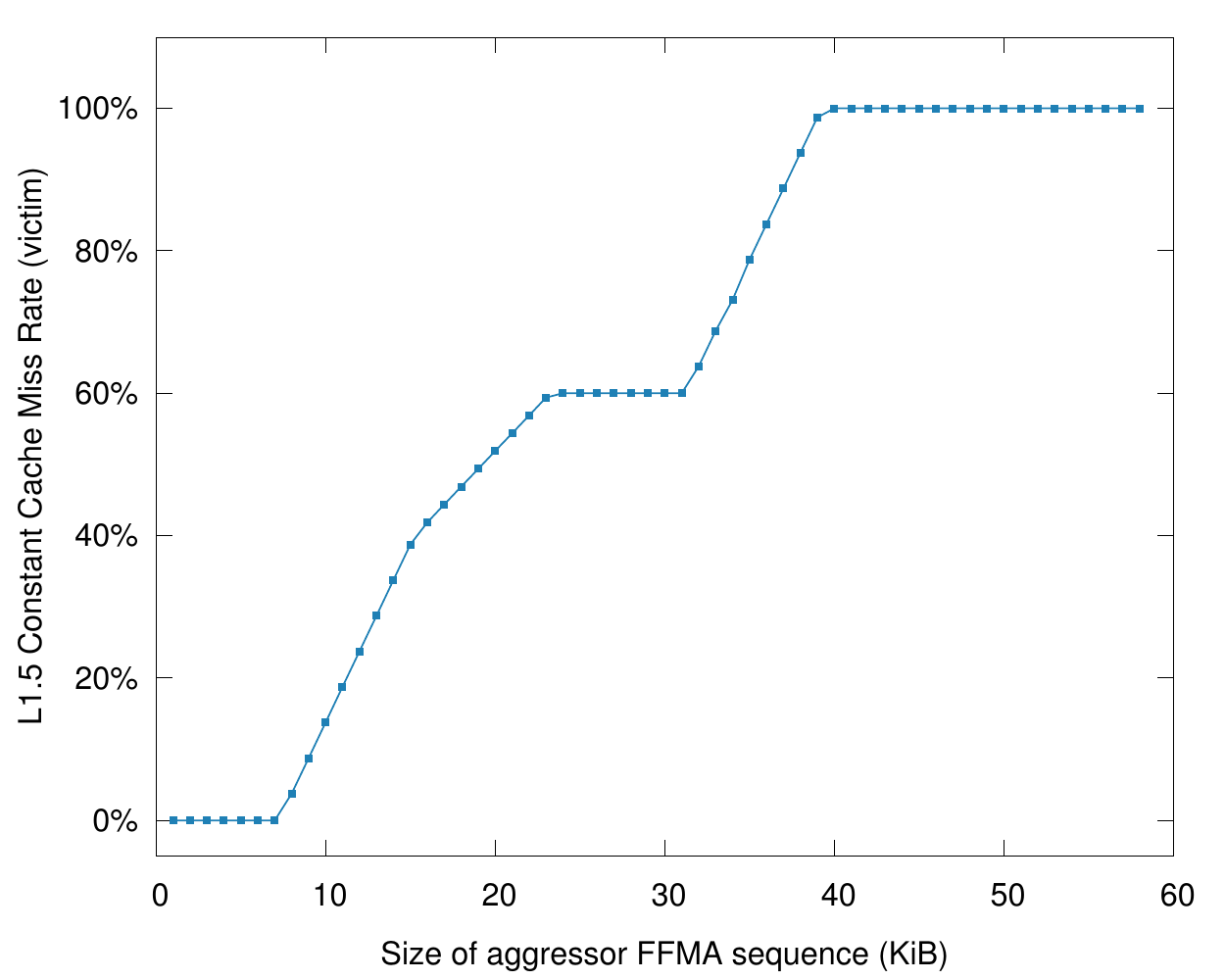}
  \caption{An aggressor-victim experiment shows that the L1.5 constant
    cache and the L1 instruction cache coincide.  We measure the miss
    rates experienced by the scan of a constant array pre-cached in
    constant L1.5 cache (victim) that follows a long sequence of
    identical \texttt{FFMA} instructions (aggressor), intentionally
    designed to cause L1 instruction cache pressure. As the
    aggressor's sequence length increases, the victim suffers
    increasing miss rates.}
  \label{fig:l15const_l1inst_shared}
\end{figure}

On Turing as in Volta, the second levels of the constant and the
instruction cache are backed by the same hardware cache.  More
precisely, the L1.5 constant cache and the L1 instruction cache
coincide.  To prove this claim, we run an aggressor-victim experiment,
in which we show that instruction sequences of increasing length
(aggressor) evict pre-populated entries in the L1.5 constant cache.
We detect these evictions by recording the execution time of a
constant array scan (victim) that we execute after the aggressor. We
use instruction sequences composed of identical \texttt{FFMA}
instructions.

Experimental results (Figure~\ref{fig:l15const_l1inst_shared}) show
that longer instruction sequences in the aggressor cause
correspondingly higher miss rates in the victim. We observed victim
miss rates vary from 0\% to 100\%.

As in previous architectures, constant memory accesses on Turing
support broadcasting (see Figure~\ref{fig:const_latency}). When all
threads within a warp access the same address, the constant memory
sends data to all threads simultaneously.  When threads visit
diverging addresses, the accesses are serialized.

\begin{figure}
  \center
  \includegraphics[width=\columnwidth]{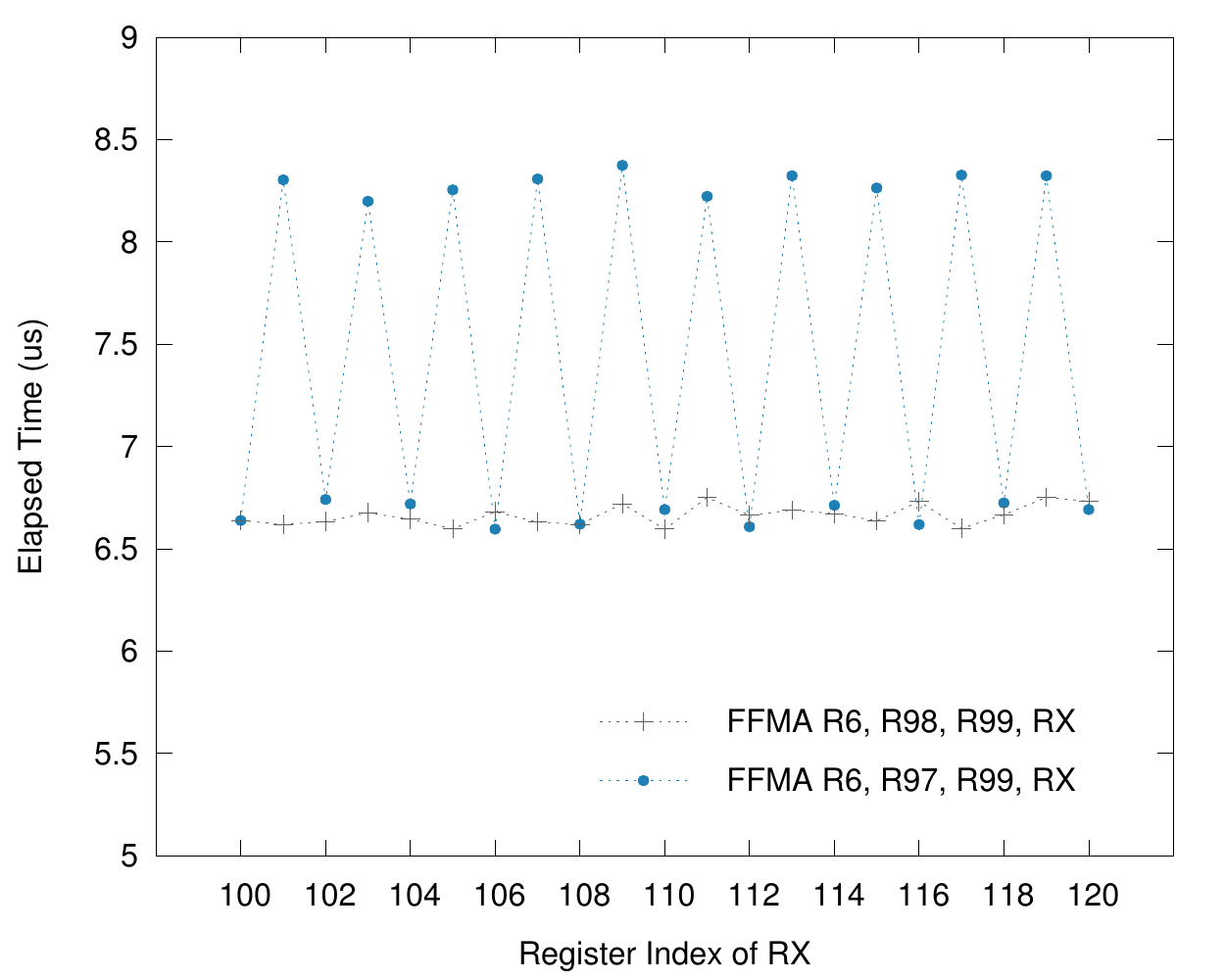}
  \caption{Register bank conflicts affect the execution time of
    instructions.  Charted is the execution time taken by long
    sequences of identical \texttt{FFMA} instructions, as we vary one
    source register (\texttt{RX}). In both sequences \texttt{R6}, the
    destination sequence, is irrelevant.  In sequence \texttt{FFMA R6,
      R97, R99, RX}, the choice of \texttt{RX} causes a conflict when
    \texttt{RX} is odd: the other two source operands are already
    using both ports from bank 1, and a third access cannot occur
    in the same clock cycle.  In sequence \texttt{FFMA R6, R98, R99,
      RX}, no choice of \texttt{RX} can cause a conflict because
    \texttt{R98} and \texttt{R99} are on different banks.}
  \label{fig:reg-bank}
\end{figure}

\section{Registers}
\label{sec:registers}

\subsection{Register File Banks}
\label{sec:registerbanks}
Turing and Volta use a physical register file of 16,384, 32-bit
elements in each processing block. Thread-visible logical registers
are allocated in increments of 8, or aggregate increments of 256 for
all 32 threads in a warp. These register files are organized in 2
banks with dual 32-bit ports each, with logical registers belonging to
the bank with the index matching their name, modulo-2. Each port can
satisfy only one 32-bit read per clock cycle, and instructions in
compiled code requiring three or more operands (such as \texttt{FFMA},
the single-precision floating-point fused multiply-and-add
instruction) will suffer a stall in execution due to a \emph{register
  bank conflict} if any three source registers' names map to either
dual-ported bank.

For example:
\begin{itemize}
\item instruction \texttt{FFMA R15, R11, R12, R13} has no conflict, since source
operands R11 and R13 can be serviced by bank 1's two ports, R12 can be
serviced by one of bank 0's ports, and destination register R15 does not
use an additional port from bank 1;
\item instruction \texttt{FFMA R18, R10, R12, R16} suffers a conflict
  because R10, R12 and R16 are all in bank 0. (The destination of R18
  is irrelevant.)
\end{itemize}

Architectures prior to Volta used 4, single-ported banks, requiring
substantially more constrained register scheduling by the compiler,
but there are opportunities for improvements even on the newest
devices.  In our technical report on Volta~\cite{zhe2018}, we
demonstrated performance increases of up to 15\% by minimizing bank
conflicts through careful register re-assignment.

Figure~\ref{fig:reg-bank} illustrates the effect of register bank
conflicts on instruction latency on the T4 GPU. We use long sequences
of identical \texttt{FFMA} instructions in which we vary one source
register index (RX) to cause conflicts. Since the T4 GPU has dual-ported
register banks, a conflict will only happen when all three 32-bit
source registers in an \texttt{FFMA} instruction belong to the same
bank. In every instruction of form \texttt{FFMA R6, R97, R99, RX} in
the benchmark, R97 and R99 are in bank 1; if RX also sits in bank 1, a
conflict will occur. (R6 is irrelevant as it is a destination
register.) In instruction sequence \texttt{FFMA R6, R98, R99, RX},
because R98 and R99 sit in different banks, there is no choice of RX
that can cause three reads from the same bank.

\subsection{Uniform Registers}

As per NVidia's documentation, Turing introduces a new feature
intended to improve the maximum achievable arithmetic throughput of
the main, floating-point capable datapaths, by adding a separate,
integer-only, scalar datapath (named the \emph{uniform datapath}) that
operates in parallel with the main datapath.

This design is intended to accelerate numerical, array-based,
compute-bound workloads that occupy the main datapaths almost
completely with floating-point instructions, typically \texttt{FFMA}
or \texttt{HMMA}, but also contain a few integer operations, typically
updating array indices, loop indices or pointers; or performing array
or loop boundary checks.

These few integer instructions spoil the instruction mix, and prevent
the main datapaths from ingesting a 100\% pure stream of \texttt{FFMA}
or \texttt{HMMA}. In these circumstances, even a small fraction of
integer instructions can hurt the overall arithmetic throughput,
lowering it significantly from its theoretical maximum.

On Turing, the compiler has the option to push these integer
operations onto the separate uniform datapath, out of the way of the
main datapath.  To do so, the compiler must emit \emph{uniform
  datapath instructions}.

Regular instructions can access both uniform and regular registers.
Uniform datapath instructions, instead, focus on uniform instructions
almost exclusively.

While at this time we have not been able to stimulate the generation
of uniform datapath instructions by the compiler, we were able to
enumerate the 64 uniform registers supported by Turing (including a
Uniform Zero Register \texttt{URZ} and 63 general-purpose uniform
registers \texttt{UR0}--\texttt{UR62}) by systematically disassembling
packed uniform instructions.

\subsection{Regular Registers}

Instructions on Turing still supports the 256 regular registers
(including the general-purpose \texttt{R0}--\texttt{R254} and the Zero
Register \texttt{RZ}).

We found that the \texttt{cuobjdump --dump-resource-usage} command
(that prints a kernel's register usage) reports a count that
includes both regular and uniform registers.  The upper limit of total
registers used in any CUDA kernel is 256, unchanged from Volta.

We confirmed this result by patching the register count in the section
header of a CUDA kernel to values above 256, and determining that
\texttt{cuobjdump} only recognizes 256 registers at most.

\begin{figure}[t]
  \center
  \includegraphics[width=\columnwidth]{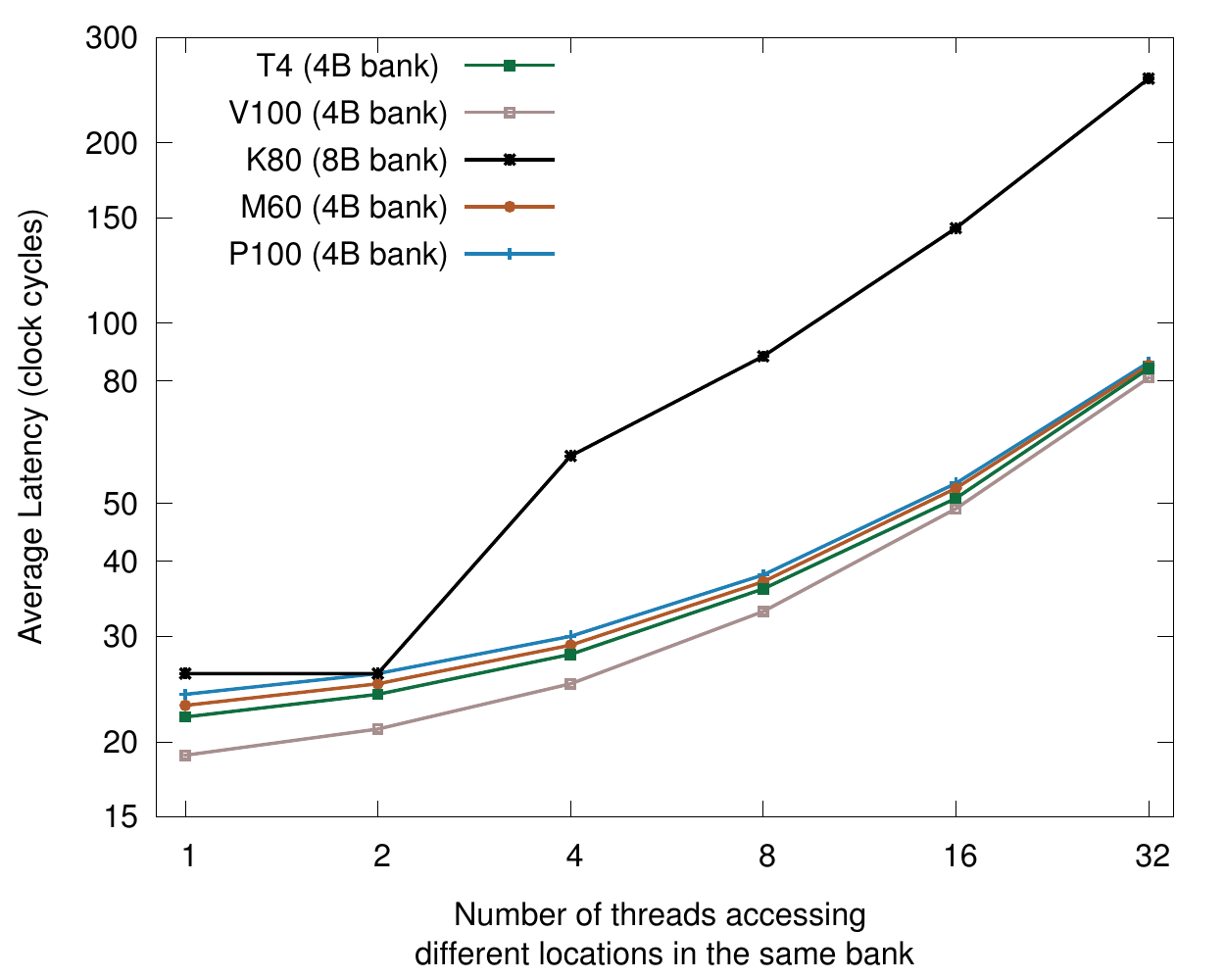}
  \caption{Shared memory latency increases under contention. Both axes
    use exponential scales. We used a stride factor multiplying the
    thread index as an offset to load data from shared memory. Each
    thread visits one 32-bit element and measures the average access
    latency. The benchmark warms shared memory before recording clock
    cycles.}
  \label{fig:shared-latency}
\end{figure}

\begin{figure}[t]
  \center
  \includegraphics[width=\columnwidth]{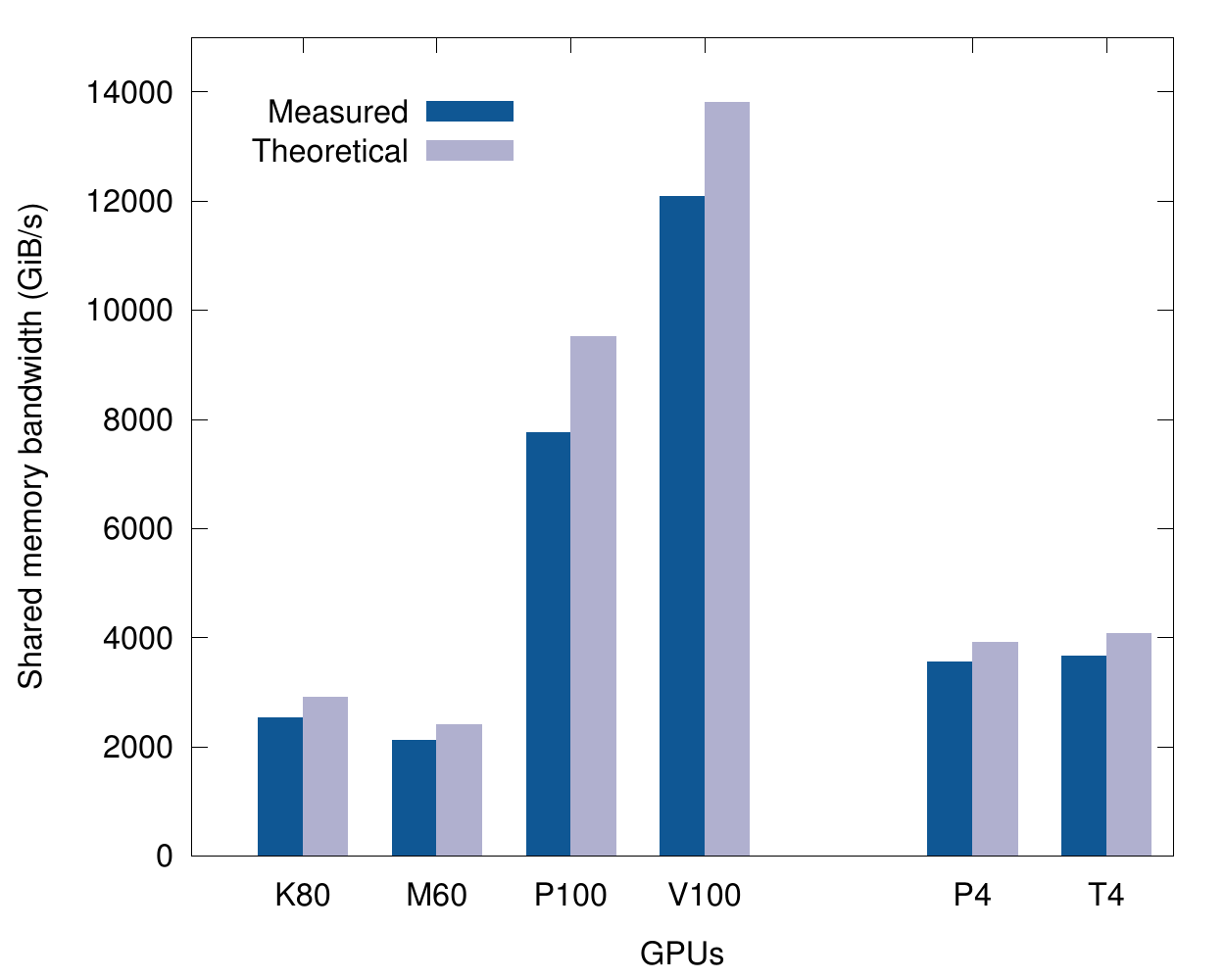}
  \caption{Theoretical and measured shared memory bandwidth on the
    considered GPUs.  The theoretical limits are given by the minimum
    of product $P \cdot B_s \cdot w_s \cdot f_g $ and
    product $P \cdot B_s \cdot n_{LSU} \cdot f_g.$
    The meaning of all factors in these products is explained in Table~\ref{tab:mem-hierarchy}.}
  \label{fig:shared-bw}
\end{figure}

\section{Shared memory}

The T4 GPU has up to 64 KiB of shared memory (configurable by the
user) that offers low latency and high memory bandwidth. In this
section, we characterize shared memory performance, including
performance under contention.

\subsection{Latency}
Turing's shared memory enjoys relatively low latency among the GPUs we
examined (Figure~\ref{fig:shared-latency}). Only the V100 GPU exhibits
lower shared memory latency than the T4 GPU.

On all GPUs except for Kepler, the measured average access latency
monotonically increases with the number of conflicts in a warp.
Kepler is the only GPU adopting dual-ported shared memory banks,
allowing any two threads to alias on any given bank without penalty
and resolving two further aliases at a time for conflicted banks.

\subsection{Bandwidth}
Due to their large number of streaming multiprocessors, the V100 and
P100 GPUs provide the highest theoretical and measured shared memory
bandwidth (Figure~\ref{fig:shared-bw}).

As benchmarking is concerned, on Kepler, Maxwell, Pascal and Volta, we
were able to rely on \texttt{nvprof} to collect shared memory metrics.
On Turing, because \texttt{nvprof} does not support shared memory
metrics collection on that GPU, we resorted to adopting the following
custom-tailored benchmark:
\begin{lstlisting}[basicstyle={\scriptsize\ttfamily}]
// Pointer-chasing shared memory bandwidth benchmark
// dData  : Pointer-chase array
// dSink  : Side-effect destination variable (prevents code elimination)
// repeat : Count of pointer-chase steps requested

// To ensure all LSUs in an SM are used, use >= 128 threads
#define THREAD_NUM 1024

// shared memory per block
#define PCHASE_SIZE 8*THREAD_NUM

__global__ void bandwidthTest(uint32_t * dData,
                              uint32_t * dSink,
                              uint32_t   repeat){
  // Pointer-chase starting position in shared memory
  uint32_t sid = threadIdx.x;

  // The pointer-chase array in shared memory
  __shared__ DTYPE shrData[PCHASE_SIZE];

  // Initialize the pointer-chase array in shared memory
  for (uint32_t i = sid; i<PCHASE_SIZE; i+=THREAD_NUM)
    shrData[i] = dData[i];

  // Synchronize threads in a same block
  __syncthreads();

  // Scan the shared-memory array with the p-chase method
  unsigned next=sid;
  for (uint32_t j = 0; j < repeat; j++) {
    next = shrData[next];
  }

  // Side effect to prevent the compiler from eliminating this code
  dSink[sid] = next;
}
\end{lstlisting}
This benchmark performs pointer-chase accesses to the shared memory
with a varying number of steps.  We invoke as many threads and blocks
as possible to provide enough pressure on load/store units.
We measured the execution time as we increased pointer-chase step count.

We cross-verified the correctness and accuracy of this benchmark by
running it on all architectures other than Turing (on which shared
memory metrics are supported) and confirming that the bandwidths it
measures match those computed from \texttt{nvprof} metrics.

\begin{figure}
  \center
  \includegraphics[width=\columnwidth]{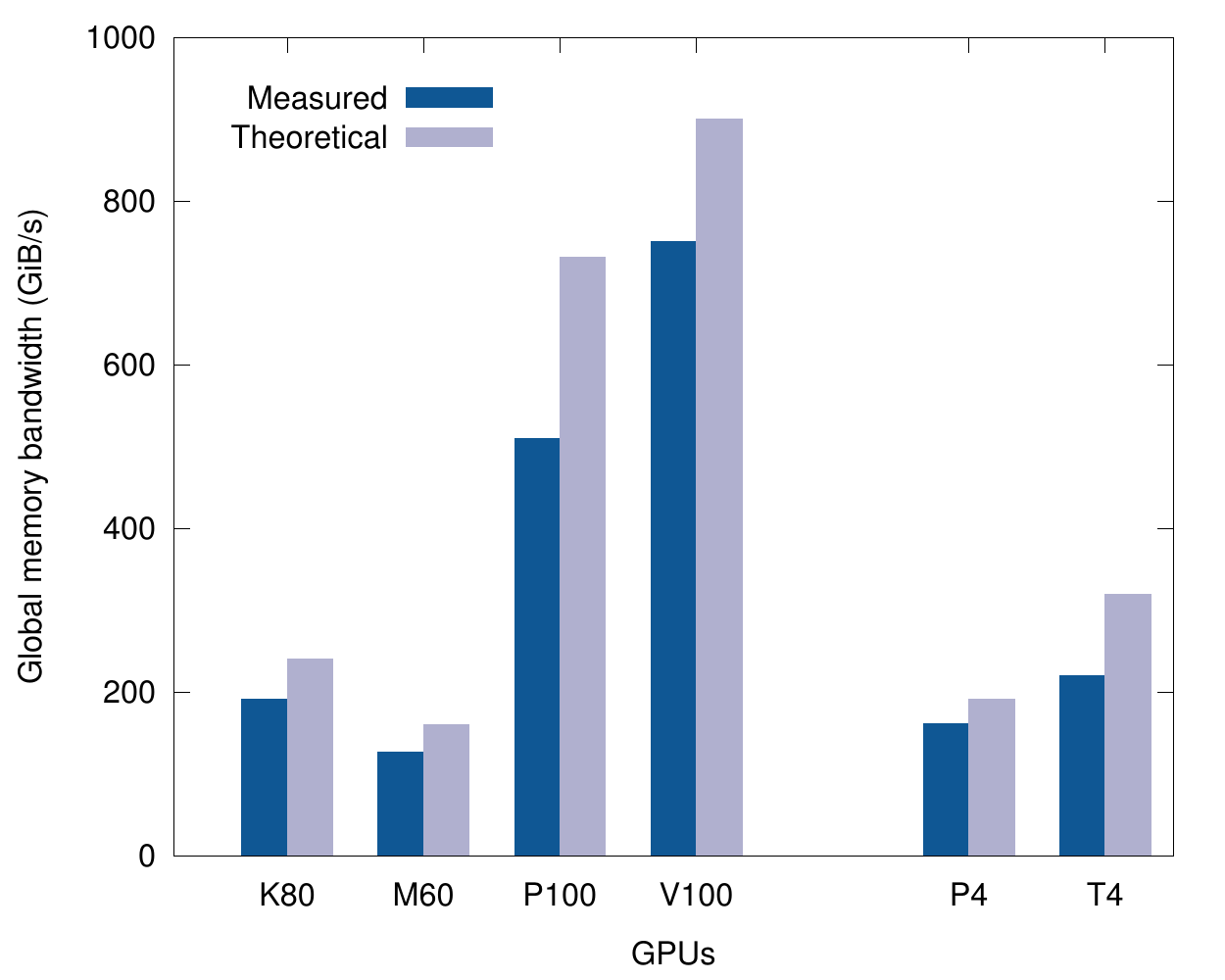}
  \caption{Theoretical and actual global memory bandwidth on all GPUs
    considered. Theoretical bounds are derived from NVidia's
    whitepapers. Actual bandwidths are the results of our benchmark,
    which loads data from a global memory array and stores it into
    another global memory array.}
  \label{fig:global-bw}
\end{figure}

\section{Global memory}
We measured the actual global memory bandwidth and compared it against
its theoretical limit for all the GPUs considered
(Figure~\ref{fig:global-bw}).

Thanks to their adoption of HBM2 memory, V100 and P100 boards
feature a significantly higher bandwidth than GPUs based on GDDR
memory. The P100 outperforms GDDR-based GPUs boards but suffers from a
large gap between actual and theoretical performance. Compared to the
P4 GPU, the T4 GPU enjoys a higher global bandwidth because of GDDR6
memory. However, the actual-to-theoretical bandwidth ratio on the T4
board is lower than on the P4 board (68.8\% vs. 84.4\%)

\begin{figure}
  \includegraphics[width=\columnwidth]{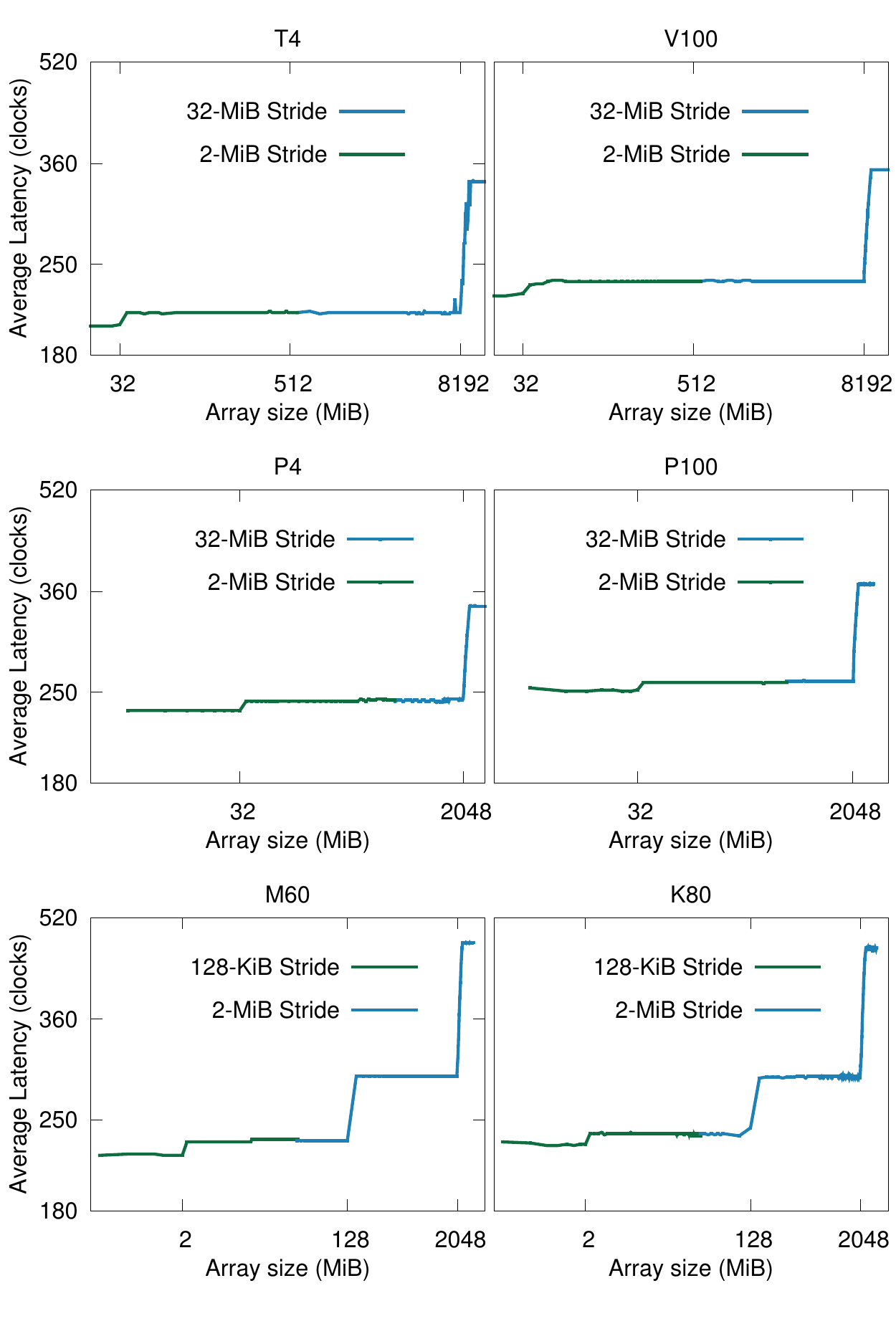}
  \caption{Global memory access latency seen by the pointer chase
    benchmark as it sweeps TLBs. The benchmarks perform a traditional
    pointer chase after a TLB warm-up scan, calculating the average
   global memory access latency with a stride of TLB page entry size.}
  \label{fig:TLB-size}
\end{figure}

\section{TLBs}

On Turing and on all other architectures that we examined, we found
that
\begin{itemize}
\item the L1 data cache is indexed by virtual addresses, and
\item the L2 data cache is indexed by physical addresses.
\end{itemize}
Because L2 is a physical cache, accesses to it involve the TLBs.  We
prove this claim by scanning a large array with L1 data cache enabled;
we size the array to exceed the L1 TLB coverage, so that accesses in
the benchmark would cause at least one level of TLB miss if L1 data
cache were indexed by physical address. As expected, we saw no TLB
misses in the second scan, as long as the stride is big enough to
cache all accesses in L1 data cache. The same benchmark shows that
addressing data in L2 data cache goes through the TLBs when the L1
data cache is disabled.

Figure~\ref{fig:TLB-size} shows that, within the available global memory size,
there are two levels of TLB on the Turing GPUs.  The L1 TLB has 2 MiB page
entries and 32 MiB coverage. The coverage of the L2 TLB is about 8192 MiB, which is
the same as Volta.

\chapter{Instruction latency and throughput}

In this chapter, we report on the latency of native Turing
instructions.  We also benchmark the performance of atomics operations
on Turing and compare it with that of older devices. We evaluate the
floating-point performance in single, double and half precision on a
T4 GPU, and evaluate the updated Tensor Cores.

\section{Native instructions}

Turing and Volta's instructions typically exhibit lower latency than
Pascal and older GPU generations, but Turing does not seem to offer
instruction latency improvements over Volta. In this section, we
report the latency of common instructions on Turing, Volta and Pascal
in Table~\ref{tab:native-latency}.

\begin{table}
  \caption{Latency of frequently used instructions on Volta and Pascal.}
  \label{tab:native-latency}
  \center
  \footnotesize
  \begin{tabular}{lll}
  \toprule
   Architecture & Instructions                                       & Latency (cycles) \\
  \midrule
   Pascal       & BFE, BFI, IADD, IADD32I, FADD, FMUL, FFMA, FMNMX,  & 6        \\
                & HADD2, HMUL2, HFMA2, IMNMX, ISCADD, LOP, LOP32I,   &          \\
                & LOP3, MOV, MOV32I, SEL, SHL, SHR, VADD, VABSDIFF,  &          \\
                & VMNMX, XMAD                                        &          \\
                &                                                    &          \\
                & DADD, DMUL, DFMA, DMNMX                            & 8        \\
                & FSET, DSET, DSETP, ISETP, FSETP                    & 12       \\
                & POPC, FLO, MUFU, F2F, F2I, I2F, I2I                &$\sim$14  \\
                & IMUL, IMAD                                         &$\sim$86  \\
  \midrule
   Volta        & IADD3, SHF, LOP3, SEL, MOV, FADD, FFMA, FMUL,      & 4        \\
                & ISETP, FSET, FSETP                                 &          \\
                & IMAD, FMNMX, DSET, DSETP                           & 5        \\
                & HADD2, HMUL2, HFMA2                                & 6        \\
                &                                                    &          \\
                & DADD, DMUL, DFMA                                   & 8        \\
                & POPC                                               &$\sim$10  \\
                & FLO, BREV, MUFU                                    &$\sim$14  \\
  \midrule
   Turing       & IADD3, SHF, LOP3, SEL, MOV, FADD, FFMA, FMUL,      & 4        \\
                & ISETP, FSET, FSETP                                 &          \\
                & IMAD, FMNMX, DSET, DSETP                           & 5        \\
                & HADD2, HMUL2, HFMA2                                & 6        \\
                &                                                    &          \\
                & POPC, FLO, BREV, MUFU                              &$\sim$15  \\
                & DADD, DMUL                                         &$\sim$48  \\
                & DFMA, DSET, DSETP                                  &$\sim$54  \\
  \bottomrule
  \end{tabular}
\end{table}

As the Turing whitepaper~\cite{tu104} mentions, the dependent-issue
latency for core FMA math operations is 4 clock cycles, the same
as on Volta.

On Turing, we found that most integer, single- and half-precision
instructions have similar latencies as those on Volta, whereas
double-precision instructions increased their latency above 40 cycles.

On Volta, most integer and single-precision instructions have a
latency of 4 cycles. In our previous work we determined that most Volta
double-precision instructions have a latency of 8 cycles, and
half-precision instructions have a latency of 6 cycles.

On Maxwell and Pascal, instructions IMAD and IMUL
have a long latency because they are emulated.

On Pascal, most integer and single-precision instructions have a
latency of 6 cycles; double-precision instructions have a latency of 8
cycles; more complex instructions, some of which run on the SFU,
require 14 cycles.

{\noindent\small\textbf{Experimental setup.} Measuring \emph{dependent issue}
  instruction latency on a software-scheduled GPU requires the use of
  custom-tailored benchmarks designed as follows. To measure the
  latency of instruction A, we add a second instruction B that depends
  on A, then set the control word that regulates A's execution:
\begin{itemize}
\item if A has fixed latency, we choose a B that consumes A's
  output. We decrease A's stall cycles in its control word, till A's
  result consumed by B is incorrect. The last stall value producing
  correct results is A's latency;
\item if A has variable latency, we choose a B of known latency, then
  set control flags to create an artificial read/write dependency
  between A and B. We let the scheduler wait for the dependency, then
  measure the pair's cumulative latency with a bracket of
  \texttt{CS2R} instructions, and obtain A's latency by subtracting
  B's known one.
\end{itemize}
}

\begin{figure}
  \includegraphics[width=\textwidth]{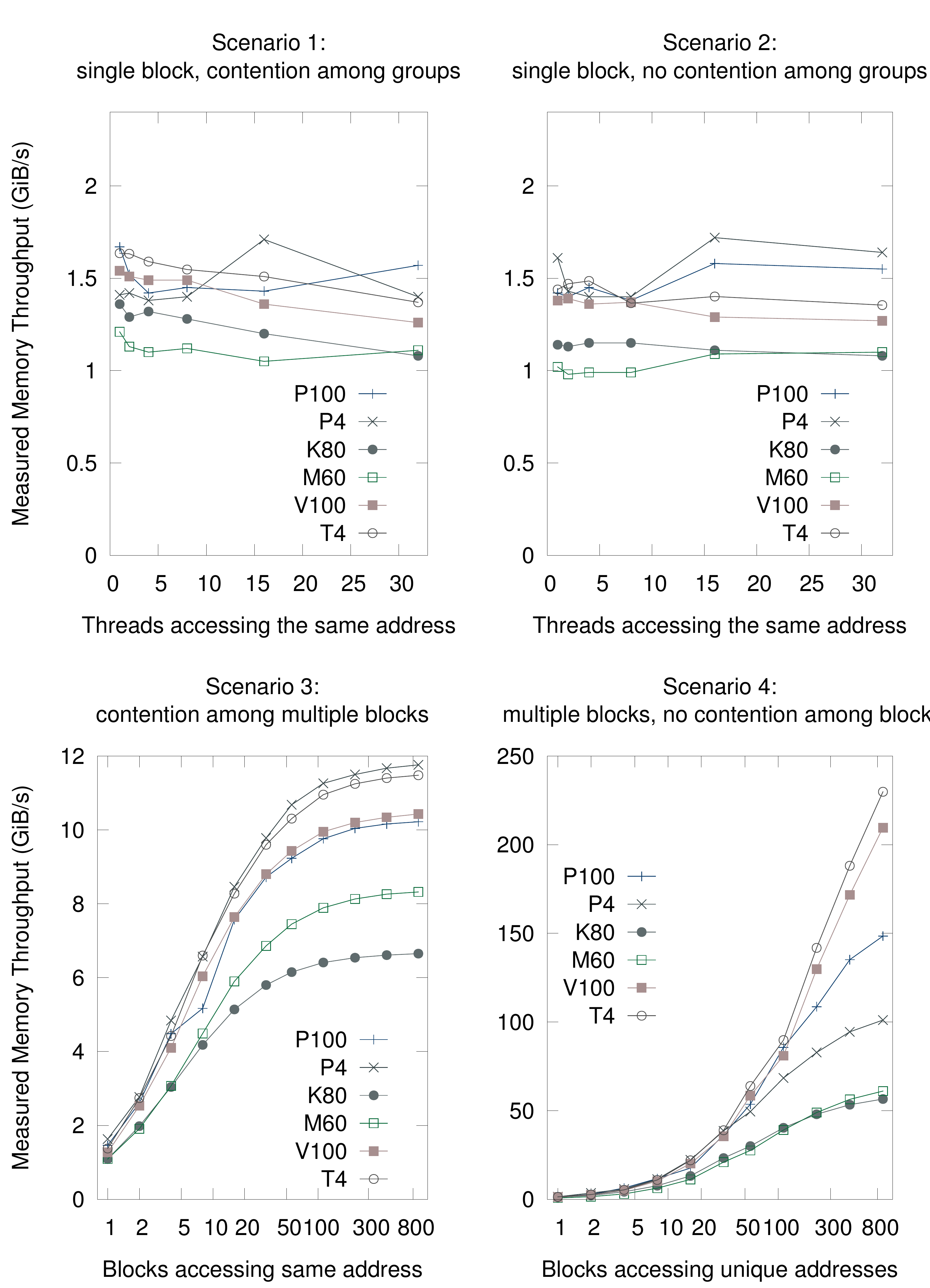}
  \caption{Throughput of \texttt{atomicAdd} operations on global
    memory, measured in four contention scenarios.}
  \label{fig:glbmem-thput}
\end{figure}

\section{Atomic operations}

Our measurements show that atomic operations on shared memory have a
slightly longer latency on Turing than on Volta, but shorter than
Pascal and older generations. In Table~\ref{tab:atomicadd-lat}, we
report those latencies expressed in clock cycles. The comparison is
meaningful even in real terms because the different GPUs adopt similar
clock frequencies (reported in Table~\ref{tab:mem-hierarchy})

As atomics on global memory are concerned, latency seems to have
increased on the T4 device compared with V100. The M60 GPU had the
best latency among all GPU considered.

Notably, Kepler is the only architecture where shared memory atomics
are slower than global memory one, and by a large margin
(4$\times$ to 8$\times$).  This is due to Kepler's lack of hardware
support for shared memory atomics. Moreover, its emulated atomics
degrade quickly under contention. Later architectures support atomics
in hardware, and offer low-latency atomics, even in presence of
contention.

\begin{table}
  \caption{Latency of atomic operations on shared and global memory, in
  clock cycles.}
  \label{tab:atomicadd-lat}
  \center
  \footnotesize
  \begin{tabular}{r|rrrrrr|rrrrrr}
    \toprule
    & \multicolumn{6}{c|}{Shared memory} & \multicolumn{6}{c}{Global memory} \\
    & & \\
    Contention     & T4    & \hspace{-2ex}V100  & \hspace{-2ex} P100  & P4 & \hspace{-1ex} M60 & \hspace{-1ex}K80 & T4    & \hspace{-2ex}V100  & \hspace{-2ex}P100  & P4    & \hspace{-1ex}M60   & \hspace{-1ex}K80 \\
  \midrule
  none             & 8     & 6     & 15    & 16    & 17    &    93    & 76    & 36    & 26    & 30    & 24    &  29 \\
  2 threads        & 10    & 7     & 17    & 18    & 19    &   214    & 72    & 31    & 31    & 50    & 26    &  69 \\
  4 threads        & 14    & 11    & 19    & 25    & 25    &   460    & 73    & 32    & 48    & 50    & 41    &  96 \\
  8 threads        & 22    & 18    & 30    & 30    & 31    &   952    & 81    & 41    & 48    & 51    & 41    & 152 \\
  16 threads       & 37    & 24    & 46    & 46    & 47    & 1,936    & 97    & 58    & 50    & 51    & 46    & 264 \\
  32 threads       & 69    & 66    & 78    & 78    & 79    & 4,257    & 116   & 76    & 50    & 51    & 46    & 488 \\
  \bottomrule
  \end{tabular}
\end{table}

We measured these latencies with benchmarks designed in the following
manner: we determine the latency of atomic instruction A by following
it with a load instruction B, of known latency, that visits the same
location. We deduce A's latency from that of pair (A,B) as described
in the previous section.

Figure~\ref{fig:glbmem-thput} reports the throughput measured on GPUs
from Kepler to Turing in presence of contention, in four scenarios:
\begin{itemize}
  \item \emph{Scenario 1}, one block of 1,024 threads. Of these, $R$ threads
    access the same address, while the others access distinct, sequential
    addresses in global memory. 8 groups of threads access the same L2 cache
    line;
  \item \emph{Scenario 2}, one block of 1,024 threads. Of these, $R$ threads
    access the same address, while the others access sequential L2 cache
    lines in global memory, with every group of threads accessing a single L2
    cache line;
  \item \emph{Scenario 3}, a variable number of blocks, of 1,024 threads each.
    All threads in all blocks access the same address; heavy contention
    exists among blocks;
  \item \emph{Scenario 4}, a variable number of blocks, of 1,024 threads each.
    All threads within a block access the same address. Different blocks
    access distinct addresses; no contention exists among blocks.
\end{itemize}

The T4 GPU doesn't achieve the highest throughput in the scenarios
with contention and the scenarios on single SM. The only scenario in
which the T4 GPU provides the best performance is on multiple SMs and
without contention among SMs. In all scenarios, from Maxwell to Pascal
the aggregate throughput increase substantially.

\begin{figure}[t]
  \center
  \includegraphics[width=\columnwidth]{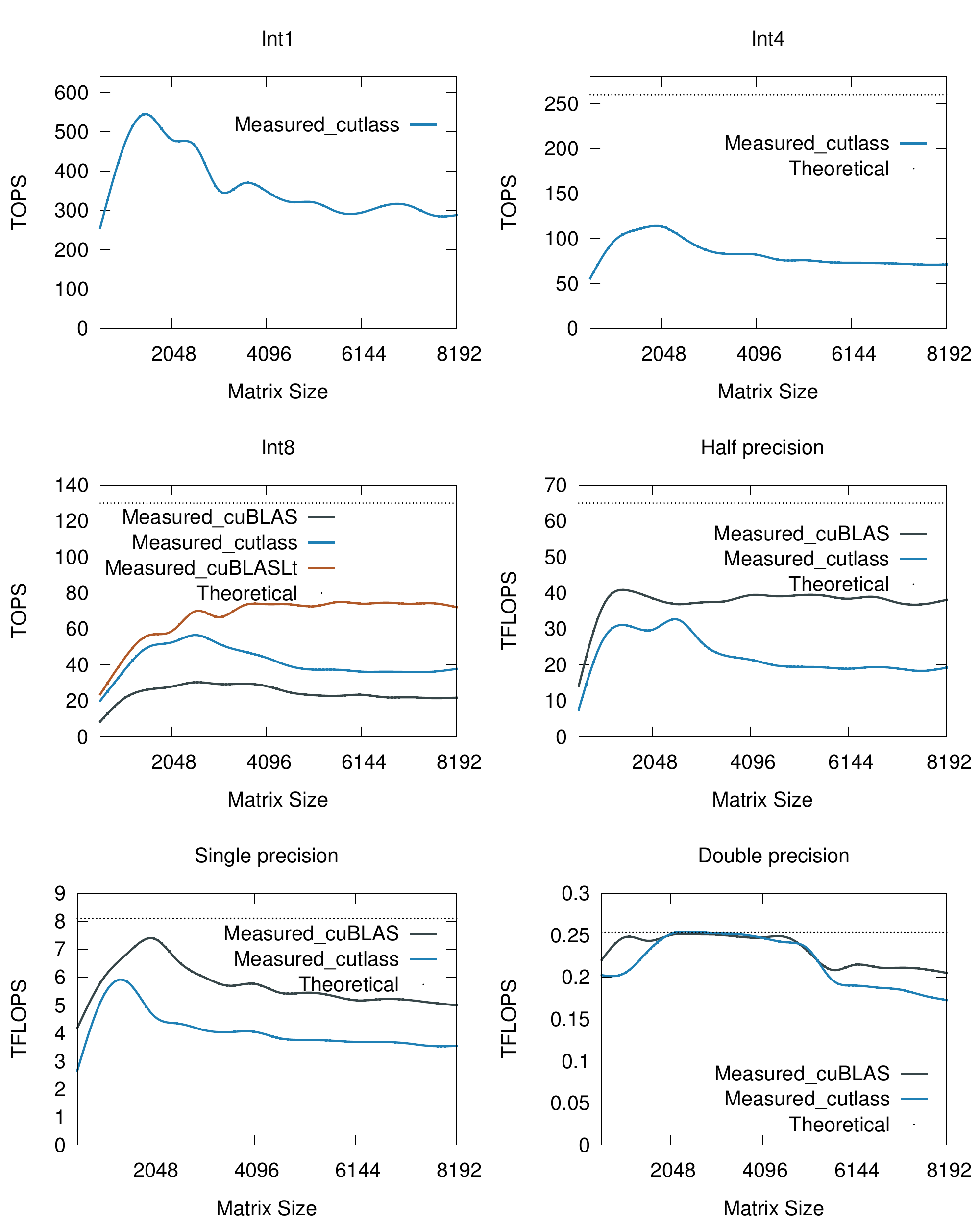}
  \caption{Floating-point performance of cuBLAS and CUTLASS matrix
    multiplication on a T4 GPU running at 1,590 MHz. }
  \label{fig:cublas-flops}
\end{figure}

\section{New Tensor Core instructions}

The Turing architecture refreshes its Tensor Cores by offering support
for a wider range of operand types than Volta.  Specifically, Tensor
Cores as introduced in Volta were designed to offer high throughput
when performing matrix math on half-precision floating point operands;
on Turing, Tensor Cores add support for short integer operands:
\texttt{int8}, \texttt{int4} and \texttt{int1}.

Moreover, Turing offers new instructions that allow to express matrix
math more succinctly. To demonstrate that, we will compare the Volta
and the Turing code generated by the compiler for the same warp-level
primitive \texttt{wmma::mma\_sync()}. Readers will recognize this
example from Chapter 4.3 of our technical report on
Volta~\cite{zhe2018}.

When targeting Volta, NVCC compiles one example invocation of the
primitive into the following 16 \texttt{HMMA.884.F32.F32.*}
instructions:
\begin{lstlisting}[basicstyle={\scriptsize\ttfamily}]
  HMMA.884.F32.F32.STEP0 R8,  R26.reuse.COL,    R16.reuse.COL,  R8  ;
  HMMA.884.F32.F32.STEP1 R10, R26.reuse.COL,    R16.reuse.COL,  R10 ;
  HMMA.884.F32.F32.STEP2 R4,  R26.reuse.COL,    R16.reuse.COL,  R4  ;
  HMMA.884.F32.F32.STEP3 R6,  R26.COL,          R16.COL,        R6  ;

  HMMA.884.F32.F32.STEP0 R8,  R20.reuse.COL,    R18.reuse.COL,  R8  ;
  HMMA.884.F32.F32.STEP1 R10, R20.reuse.COL,    R18.reuse.COL,  R10 ;
  HMMA.884.F32.F32.STEP2 R4,  R20.reuse.COL,    R18.reuse.COL,  R4  ;
  HMMA.884.F32.F32.STEP3 R6,  R20.COL,          R18.COL,        R6  ;

  HMMA.884.F32.F32.STEP0 R8,  R22.reuse.COL,    R12.reuse.COL,  R8  ;
  HMMA.884.F32.F32.STEP1 R10, R22.reuse.COL,    R12.reuse.COL,  R10 ;
  HMMA.884.F32.F32.STEP2 R4,  R22.reuse.COL,    R12.reuse.COL,  R4  ;
  HMMA.884.F32.F32.STEP3 R6,  R22.COL,          R12.COL,        R6  ;

  HMMA.884.F32.F32.STEP0 R8,  R2.reuse.COL,     R14.reuse.COL,  R8  ;
  HMMA.884.F32.F32.STEP1 R10, R2.reuse.COL,     R14.reuse.COL,  R10 ;
  HMMA.884.F32.F32.STEP2 R4,  R2.reuse.COL,     R14.reuse.COL,  R4  ;
  HMMA.884.F32.F32.STEP3 R6,  R2.COL,           R14.COL,        R6  ;
\end{lstlisting}

\noindent When targeting Turing, NVCC compiles the same primitive
invocation into only 4 \texttt{HMMA} instructions of a new kind, that
contain the new \texttt{.1688} infix:
\begin{lstlisting}[basicstyle={\scriptsize\ttfamily}]
  # Turing rendition
  HMMA.1688.F32 R8,  R12, R22, R8 ;
  HMMA.1688.F32 R4,  R12, R23, R4 ;
  HMMA.1688.F32 R8,  R2,  R24, R8 ;
  HMMA.1688.F32 R4,  R2,  R25, R4 ;
\end{lstlisting}

\section{Arithmetic performance}

We evaluated arithmetic performance by benchmarking matrix-matrix
multiplications using functions from the cuBLAS 10.1 library and
template functions from cutlass 1.2, on integer operands and
floating-point ones of different precisions.  We report arithmetic
throughput in TOPS and TFLOPS, when operating on integer and
floating-point values respectively. In all experiments, the T4 GPU was
running at a clock frequency of 1,590 MHz.

In half, single and double precision, cuBLAS provides higher
arithmetic throughput than cutlass. This is because the cuBLAS library
has been specifically optimized for the Turing architecture. For
\texttt{int8} precision, two APIs are available in cuBLAS 10.1:
\begin{itemize}
  \item BLAS-like extension function \texttt{cublasGemmEx}, which
    invokes native CUDA core implementations, and
  \item the new light-weight \texttt{cublasLtMatmul} function, which
    supports \texttt{int8} native TensorCore implementations.
    \end{itemize}
For \texttt{int8}, the throughput of (\texttt{cublasLtMatmul}) is much
higher than the throughput of (\texttt{cublasGemmEx}).  At the time of
this writing, only cutlass supports \texttt{int4} and \texttt{int1}
matrix multiplication on NVidia GPUs.

Except in double precision, benchmarks don't achieve near-peak
performance. For \texttt{int8} and \texttt{int4}, cutlass
implementations don't achieve 50\% of theoretical throughput on the T4
GPU (Figure~\ref{fig:cublas-flops}).

In Table~\ref{tab:matmul-throughput} we compare the arithmetic
throughputs achieved on T4 and P4 GPUs on matrix multiplication at
different precisions, with both boards running at the respective top
frequencies (1,590 and 1,531 MHz). The T4 GPU enjoys a higher
throughput in half precision and \texttt{int8} precision, thanks to
Tensor Cores usage.

Because the T4 and the P4 GPU have the same number of CUDA cores, we
measure similar arithmetic throughput in matrix multiplication on the
two boards, in double and single precision.  Note that
double-precision performance is hampered by the small number of native
FP64 cores available (only two per SM), as both architectures are
optimized for inference, where lower precision is more frequently
employed.

\begin{table}
  \caption{Arithmetic throughput of matrix multiplication on
    inference-oriented GPUs on floating point and integer types.}
  \label{tab:matmul-throughput}
  \center
  \small
  \begin{tabular}{lrrl }
  \toprule
                   &    T4      &    P4  & \\
  \midrule
  Double precision &     253    &    231 & GFLOPS \\
  Single precision &   7,174    &  6,944 & GFLOPS \\
  Half   precision &  41,616    &  6,571 & GFLOPS \\
  Int8   precision &  74,934    & 24,172 & GOPS  \\
  Int4   precision & 114,384    & -      & GOPS  \\
  Int1   precision & 552,230    & -      & GOPS  \\
  \bottomrule
  \end{tabular}
\end{table}

\begin{figure}[t]
  \includegraphics[width=\columnwidth]{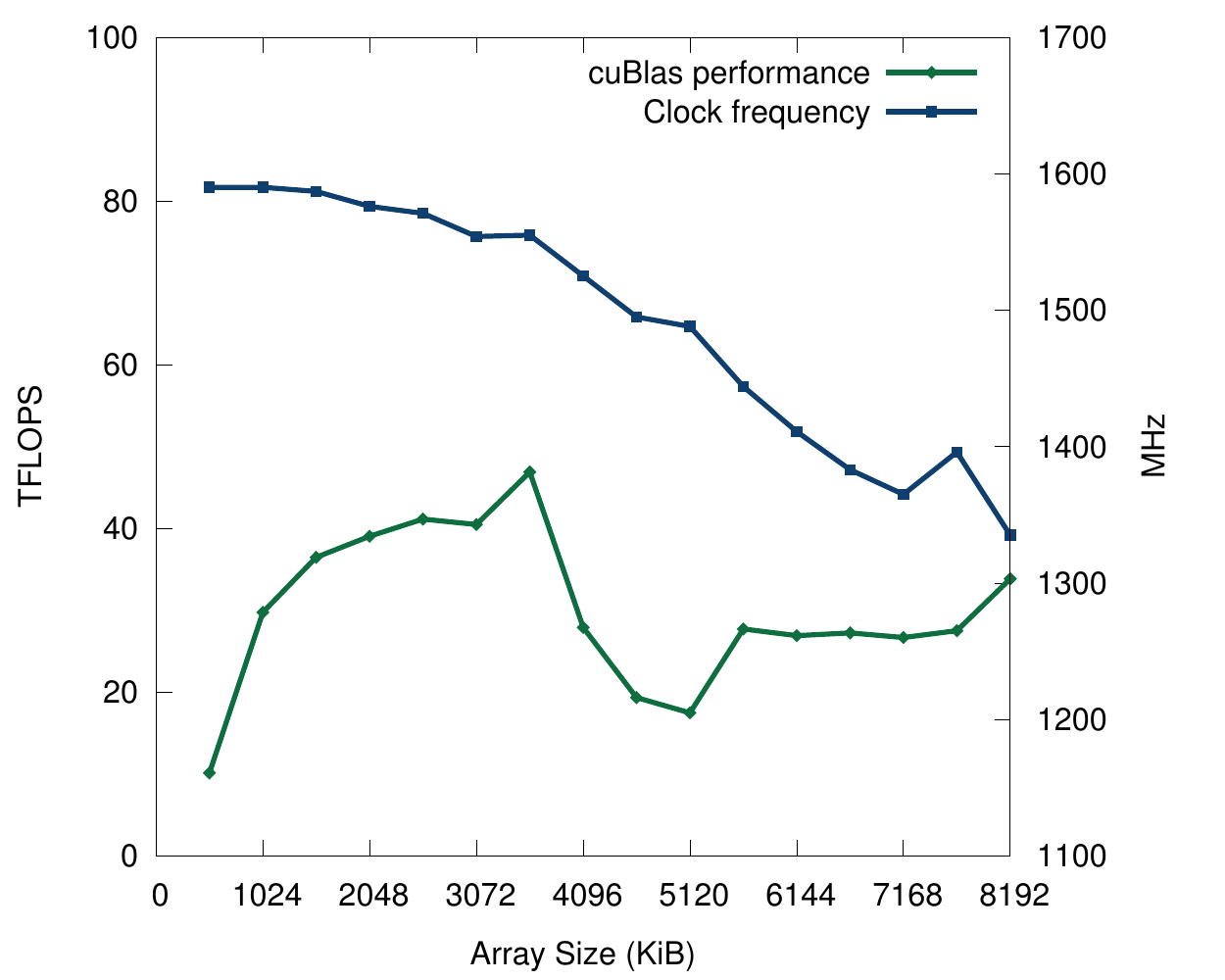}
  \caption{Clock frequency observed on a T4 GPU while continuously
    computing cuBLAS matrix multiplication. The application clock
    frequency is set to 1,590 MHz.}
  \label{fig:cublas-clock}
\end{figure}

\begin{figure}[t]
  \center
  \includegraphics[width=\columnwidth]{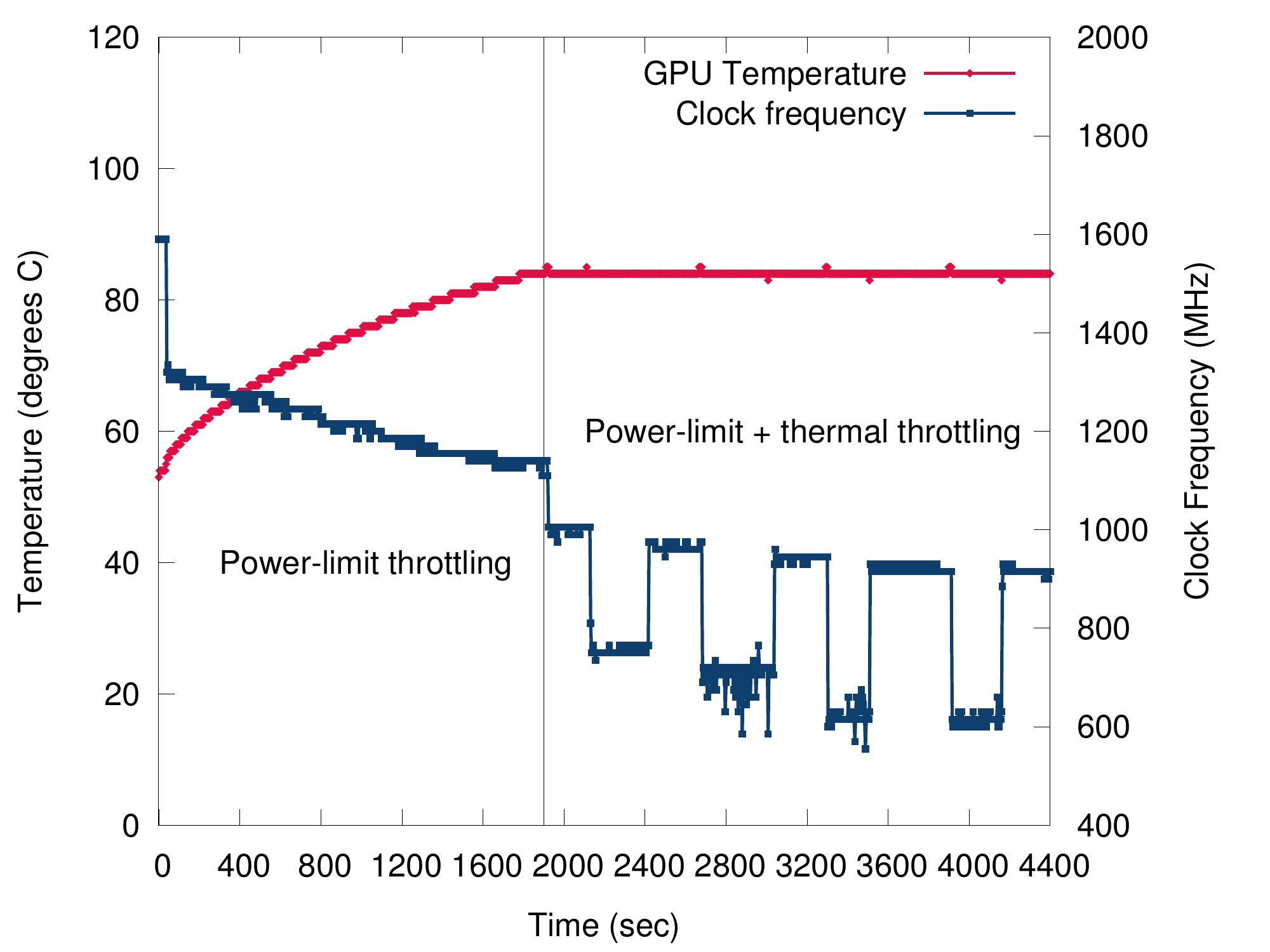}
  \caption{Temperature and clock frequency of the T4 card when
    computing a \texttt{cublasSgemm} repeatedly. The
    application clock frequency is set to 1,590 MHz.}
  \label{fig:cublas-temp-clock}
\end{figure}

\section{Performance throttling}

Most GPUs include forms of clock throttling and/or power-state
throttling to prevent exceeding either the power or thermal envelopes
if the workload is particularly demanding or the heat dissipation is
insufficient.

Our experiments show that the small form-factor T4 and P4 boards,
designed for inference applications, achieve a significantly higher
frequency-per-Watt rating than their full-size counterparts. At the
same time, they are more prone to clock throttling than their full-size
counterparts (K80, P100, V100, and M60) because of
\begin{itemize}
  \item their smaller size, which limits their heat sinks' heat
    transfer rate, and
  \item their maximum power limits set by the manufacturer, which is
    significantly lower (70W) on low-power, small form-factor boards
    than on full-size boards (250W).
\end{itemize}

{\small\noindent \textbf{Experimental setup}: All GPU specimens we
  examined adopt passive cooling. Our K80, P100, V100 and M60 experiments
  ran on Dell PowerEdge C4130 servers, which are Tesla-qualified.  Our
  T4 and P4 experiments ran on HPE Proliant DL360 Gen9 servers. This
  server model does not appear in NVidia's Tesla-qualified server
  catalog. Power and thermal performance of a GPU also depend on the
  server that hosts it, and could be suboptimal on a non-qualified
  server.  The server generation immediately following the one we
  employed (HPE Proliant DL360 Gen10) \emph{is} Tesla-qualified, but
  we were unable to arrange for an upgrade before the publication of
  this manuscript.}

In our experiments, we were able to trigger clock throttling on the T4
GPU consistently, using benchmarks based on cuBLAS matrix
multiplication kernels \texttt{cublas<t>gemm}.  On the T4 GPU, clock
throttling triggers for two reasons:
\begin{itemize}
  \item \textbf{power-limit throttling:} instantaneous power exceeds
    the power limit set by the manufacturer (70W on the T4 GPU);
  \item \textbf{thermal throttling:} the GPU reaches its maximum
    operating temperature (85\degree C on the T4 card).
\end{itemize}
Compared to power limit throttling, thermal throttling causes a more
severe clock frequency reduction.

\begin{figure}
  \center
  \includegraphics[width=\columnwidth]{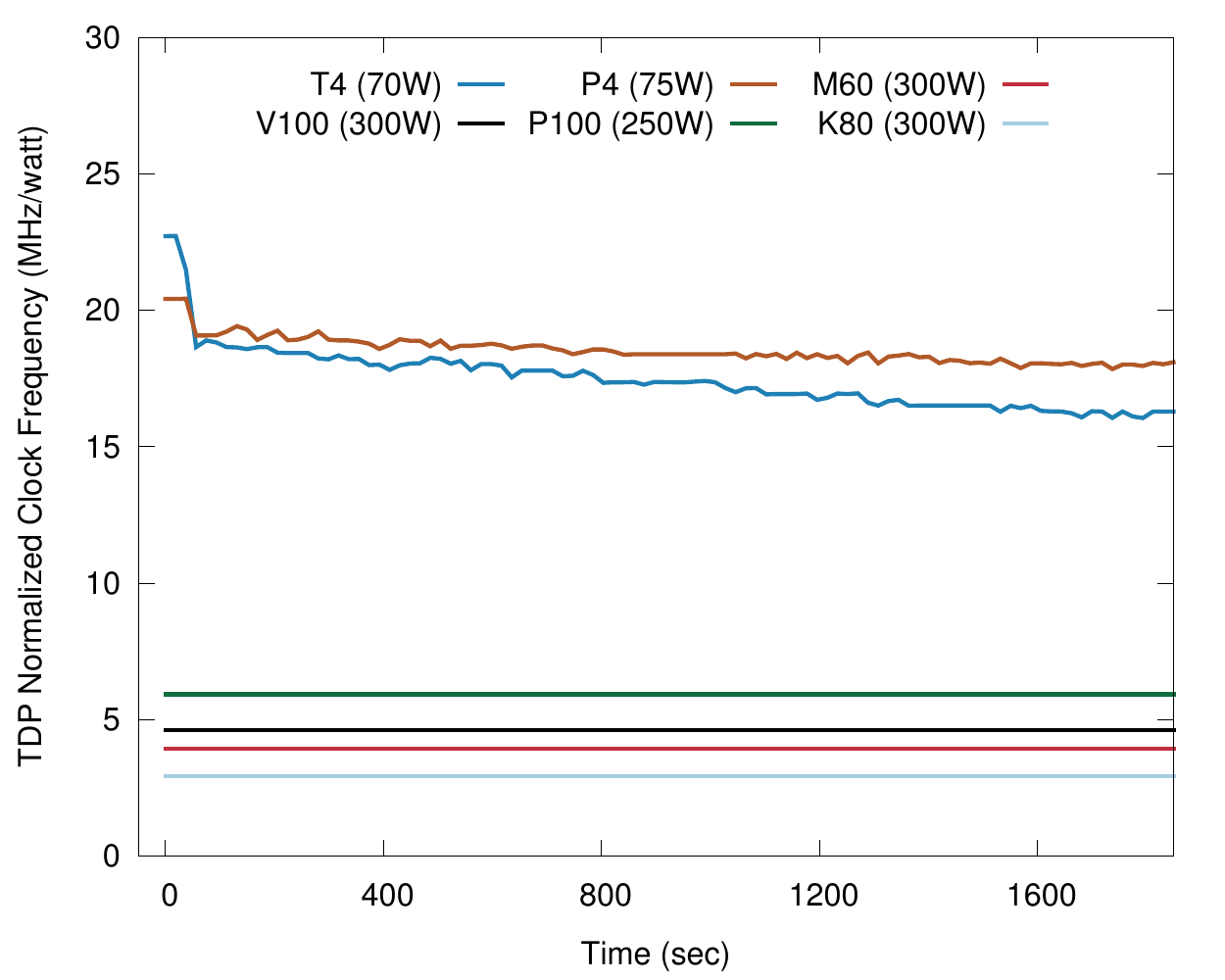}
  \caption{Clock frequency normalized to thermal design power (TDP) of
    all considered GPUs when computing an identical \texttt{cublas<t>sgemm}
    function on 1024$\times$1024 matrices repeatedly. On every GPU, we
    set the application clock frequency to its maximum supported value.}
  \label{fig:cublas-time-clock}
\end{figure}

\subsection{Power-limit throttling}
On the T4 and P4 GPUs, we saw power-limit throttling trigger very
early in our cuBLAS-based matrix multiplication experiments. On the
other hand, the V100, P100, M60 and K80 GPUs barely experienced any
power-limit throttling, due to the larger margin between actual power
consumption and its limit.

To confirm the cause of throttling, we designed an experiment that
invokes \texttt{cuBLAS<t>gemm} kernels with input matrices of growing
size. We observed the T4 GPU exceeded its power limit more and more
frequently, and lower its clock rates more and more, with growing
input sizes. The reduced clock frequency eventually hurts overall
arithmetic throughput. See Figure~\ref{fig:cublas-clock}.

In the experiment, we set the application clock for graphics on the T4
card to 1,590 MHz, and prevent GPU temperatures from exceeding the
maximum operating temperature of the T4 GPU. We record the clock
frequency of the T4 card while computing \texttt{cublas<t>gemm} in
half precision.

\subsection{Thermal throttling}

We characterized thermal throttling with a benchmark that repeatedly
launches a \texttt{cublas<t>gemm} kernel on a large matrix.  We
observed that below 85 degrees C (the maximum operating temperature),
power limit throttling causes the T4 GPU to reduce its graphics clock
with the growth of temperature.  As soon as the temperature reaches 85
degrees C, thermal throttling triggers in addition to power-limit
throttling, causing a more dramatic clock frequency step-down,
depicted in Figure~\ref{fig:cublas-temp-clock}.

\subsection{Power-limit throttling across GPU devices}

We compared the power-limit throttling behavior of the different GPUs,
by recording graphics clock over time while all cards computed endless
repetitions of the same \texttt{cublasSgemm} kernel on
1024$\times$1024 input matrices.

We noticed substantial differences between low-power GPUs (e.g., T4
and P4) and the full form-factor GPUs (K80, M60, P100, V100).  We
observe clock throttling only on the T4 and the P4 GPUs. Both cards
are only able to run at their highest supported clock frequency for a
few seconds at the very beginning of the experiment.  As temperatures
increased, clock throttling intervened and clock frequency decreased
(Figure~\ref{fig:cublas-time-clock}).

On full-height, full-length GPUs, we could not raise power consumption
enough to approach the limits and trigger throttling.

{\noindent\small\textbf{Experimental setup}: in all experiments, we set all
  graphics clocks to the highest supported value for each device. We
  turned off the AutoBoost features wherever available. We also
  ensured that only power-limit throttling was active.}

\chapter{Conclusions}

We refreshed our microbenchmark-based architectural discovery study,
updating it for the Turing architecture.  We revealed Turing's
architectural details, and compared them with previous NVidia
architectures.

We emphasize the comparison between the T4 and the P4 GPUs: both are
low-power, small-form-factor boards that target inference
applications. The T4 is based on the Turing architecture; the P4, its
predecessor, is based on Pascal.

We find that Turing uses the same instruction encoding as Volta, but
it extends Volta's instruction set; it also introduces a new register
type (uniform registers) and supports more operand types on Tensor
Cores. The new instructions allow the \texttt{nvcc} compiler to render
matrix math in fewer instructions on Turing than on Volta.

The T4 GPU also delivers a significantly higher arithmetic throughput
than the P4 on reduced-precision operands.

Turing's memory hierarchy is similar to that of Volta, with different
sizes at certain cache levels. We provided an exhaustive examination
of the differences. When compared in terms of instruction encoding,
memory hierarchy, and behavior of their processing units, the Turing
and Volta generations display continuity, and together represent a
significant departure from the Kepler and older generations.

Turing continues a trend of growth in the scheduler-to-cores ratio,
which grew from 1:48 in Kepler to 1:16 in Turing.  This trend
correlates with a growth in instruction throughput.  With their newly
introduced L0 instruction cache, Turing and Volta mitigate the penalty
associated with their longer instructions. The improved L1 data cache
offers lower latency and higher bandwidth. Their new replacement
policy also reduces cache miss rates when not using shared memory, and
the change from 4, single-ported register banks in Pascal to 2,
dual-ported banks facilitates the prevention of bank conflicts.

Compared to the Pascal P4 GPU, the Turing T4 GPU provides higher
bandwidth on L1 cache and global memory. The T4 GPU has higher
arithmetic throughput for half-precision, \texttt{int8} and
\texttt{int4} matrix multiplication thanks to its improved Tensor
Cores.  In single and double precision, the T4 and the P4 GPUs exhibit
comparable performance because they contain the same number of cores
and are clocked at similar frequencies.

Using our findings on the instruction set encoding, software designers
can optimize their code at the binary level and even construct
customized SASS assemblers able to target Turing and possibly generate
more tightly scheduled code, that delivers higher performance. Thanks
to the memory hierarchy information we disclose, developers can also
optimize their code by selecting working sets that match the cache
memories at every suitable level, thus reducing miss rates and
improving overall performance.

\chapter*{Appendix}

In this appendix, we provide the opcodes for common instructions, as
encoded in Turing's instruction encoding and, for comparison, in
Pascal's and Volta's encoding.

{
  \footnotesize
  \begin{longtable}{L{14ex}L{16ex}R{16ex}R{16ex}}
  \label{tab:pascal-opcodes}
  \endfirsthead
  \toprule
  \endhead
  \bottomrule
  \endfoot
  \endlastfoot
      \toprule
      \multicolumn{4}{c}{Floating point instructions} \\[4pt]
      Instruction &  \multicolumn{1}{c}{Pascal} & \multicolumn{1}{c}{Volta} & \multicolumn{1}{c}{Turing} \\
      \midrule
      FADD   & 0101~1100~0101~1 & 10~0010~0001   & ~10~0010~0001  \\
             & 0100~1100~0101~1 &                              &    \\
             & 0011~1001~0101~1 &                              &    \\
      FCHK   & 0101~1100~1000~1 & 011~0000~0010  & ~011~0000~0010  \\
             & 0100~1100~1000~1 &                              &    \\
             & 0011~1001~1000~1 &                              &    \\
      FCMP   & 0101~1011~1010   & --             & -- \\
             & 0101~0011~1010   & & \\
             & 0100~1011~1010   & & \\
             & 0011~0111~1010   & & \\
      FFMA   & 0101~1001~1      & 10~0010~0011   & 10~0010~0011  \\
             & 0101~0001~1      & & \\
             & 0100~1001~1      & & \\
             & 0011~0011~1      & & \\
             & 0011~0010~1      & & \\
      FMNMX  & 0101~1100~0110~0 & 010~0000~1001    & 010~0000~1001  \\
             & 0100~1100~0110~0 & & \\
             & 0011~1001~0110~0 & & \\
             & 0011~1000~0110~0 & & \\
      FMUL   & 0101~1100~0110~1 & 010~0010~0000    & 010~0010~0000  \\
             & 0100~1100~0110~1 & & \\
             & 0011~1001~0110~1 & & \\
             & 0011~1000~0110~1 & & \\
      FSET   & 0101~1000~       & 010~0000~1010    & 010~0000~1010 \\
             & 0100~1000~       & & \\
             & 0011~0001~       & & \\
      FSETP  & 0101~1011~1011   & 010~0000~1011    & 010~0000~1011  \\
             & 0100~1011~1011   & & \\
             & 0011~0111~1011   & & \\
             & 0011~0110~1011   & & \\
      FSWZADD& 0101~0000~1111~1 & 0~1000~0010~0010 & 0~1000~0010~0010  \\
      MUFU   & 0101~0000~1000~0 &    011~0000~1000 &    011~0000~1000  \\
      RRO    & 0101~1100~1001~0 & -- &   \\
             & 0100~1100~1001~0 & & \\
      \multicolumn{4}{c}{Floating point instructions (continued)} \\[4pt]
      Instruction &  \multicolumn{1}{c}{Pascal} & \multicolumn{1}{c}{Volta} & \multicolumn{1}{c}{Turing} \\
      \midrule

             & 0011~1001~1001~0 & & \\
             & 0011~1000~1001~0 & & \\
      DADD   & 0101~1100~0111~0 & 10~0010~1001   & ~10~0010~1001 \\
             & 0100~1100~0111~0 & & \\
             & 0011~1001~0111~0 & & \\
             & 0011~1000~0111~0 & & \\
      DFMA   & 0101~1011~0111   & 10~0010~1011   & ~10~0010~1011 \\
             & 0101~0011~0111   & & \\
             & 0100~1011~0111   & & \\
             & 0011~0111~0111   & & \\
             & 0011~0110~0111   & & \\
      DMNMX  & 0101~1100~0101~0 & --                           &  -- \\
             & 0100~1100~0101~0 & & \\
             & 0011~1001~0101~0 & & \\
             & 0011~1000~0101~0 & & \\
      DMUL   & 0101~1100~1000~0 & 010~0010~1000  & ~010~0010~1000  \\
             & 0100~1100~1000~0 & & \\
             & 0011~1001~1000~0 & & \\
             & 0011~1000~1000~0 & & \\
      DSET   & 0101~1001~0      & --                            & --  \\
             & 0100~1001~0      & & \\
             & 0011~0011~0      & & \\
             & 0011~0010~0      & & \\
      DSETP  & 0101~1011~1000   & 10~0010~1010   & 10~0010~1010  \\
             & 0100~1011~1000   &                              &    \\
             & 0011~0111~1000   &                              &    \\
             & 0011~0110~1000   &                              &    \\
      HADD2  & --               & 10~0011~0000   & 10~0011~0000  \\
      HFMA2  & --               & 10~0011~0001   & 10~0011~0001  \\
      HMMA   & --               & 0~0010~0011~0110  & 0~0010~0011~0110  \\
      HMUL2  & --               & 010~0011~0010  & 010~0011~0010  \\
      HSETP2 & --               & 10~0011~0100   & 10~0011~0100  \\
      HSET2  & --               & 10~0011~0011   & 10~0011~0011  \\
      FSEL   & --               & 010~0000~1000  & 010~0000~1000  \\

      \midrule
             & &     &    \\
             & &     &    \\
      \multicolumn{4}{c}{Integer Instructions} \\[4pt]
      Instruction &  \multicolumn{1}{c}{Pascal} & \multicolumn{1}{c}{Volta} & \multicolumn{1}{c}{Turing} \\
      \midrule
      BFE    & 0101~1100~0000~0 & --  & --  \\
             & 0100~1100~0000~0 &                        &    \\
             & 0011~1001~0000~0 &                        &    \\
             & 0011~1000~0000~0 &                        &    \\
      BFI    & 0101~1011~1111~0 & --  & --  \\
             & 0101~0011~1111~0 &                        &    \\
             & 0100~1011~1111~0 &                        &    \\
             & 0011~0111~1111~0 &                        &    \\
             & 0011~0110~1111~0 &                        &    \\
      FLO    & 0101~1100~0011~0 & 011~0000~0000  & 011~0000~0000  \\
             & 0100~1100~0011~0 &                              &    \\
             & 0011~1001~0011~0 &                              &    \\
             & 0011~1000~0011~0 &                              &    \\
             &                  &                              &    \\
      \multicolumn{4}{c}{Integer Instructions (continued)} \\[4pt]
      Instruction &  \multicolumn{1}{c}{Pascal} & \multicolumn{1}{c}{Volta} & \multicolumn{1}{c}{Turing} \\
      \midrule
      IADD   & 0101~1100~0001~0 & --   & --  \\
             & 0100~1100~0001~0 &                              &    \\
             & 0101~1100~0001~0 &                              &    \\
             & 0101~1101~0001~0 &                              &    \\
      IADD3  & 0101~1100~1100   & 010~0001~0000  & ~010~0001~0000  \\
             & 0100~1100~1100   &                              &    \\
             & 0011~1001~1100   &                              &    \\
             & 0011~1000~1100   &                              &    \\
      ICMP   & 0101~1011~0100   & --   & --  \\
             & 0101~0011~0100   &                              &    \\
             & 0100~1011~0100   &                              &    \\
             & 0011~0111~0100   &                              &    \\
             & 0011~0110~0100   &                              &    \\
      IMAD   & 0101~1010~0      &  10~0010~0100  & ~10~0010~0100  \\
             & 0101~0010~0      &  10~0010~0101  & ~10~0010~0101   \\
             & 0100~1010~0      &                              &    \\
             & 0011~0100~0      &                              &    \\
      IMADSP & 0101~1010~1      & --   & --  \\
             & 0101~0010~1      &                              &    \\
             & 0100~1010~1      &                              &    \\
             & 0011~0101~1      &                              &    \\
             & 0011~0100~1      &                              &    \\
      IMNMX  & 0101~1100~0010~0 & --   &  -- \\
             & 0100~1100~0010~0 &                              &    \\
             & 0011~1001~0010~0 &                              &    \\
             & 0011~1000~0010~0 &                              &    \\
      IMUL   & 0011~1000~0011~1 & ?                             & ?  \\
             & 0100~1100~0011~1 &                              &    \\
             & 0011~1001~0011~1 &                              &    \\
             & 0011~1000~0011~1 &                              &    \\
      ISCADD & 0101~1100~0001~1 & --   &  -- \\
             & 0100~1100~0001~1 &                              &    \\
             & 0011~1001~0001~1 &                              &    \\
             & 0011~1000~0001~1 &                              &    \\
      ISET   & 0101~1011~0101   & --   & --   \\
             & 0100~1011~0101   &                              &    \\
             & 0011~0111~0101   &                              &    \\
             & 0011~0110~0101   &                              &    \\
      ISETP  & 0011~0111~0110   & 010~0000~1100  & ~010~0000~1100  \\
             & 0100~1011~0110   &                              &    \\
             & 0011~0111~0110   &                              &    \\
             & 0011~0110~0110   &                              &    \\
      LEA    & 0101~1011~1101~0 & 010~0001~0001  & ~010~0001~0001  \\
             & 0101~1011~1101~1 &                              &    \\
             & 0100~1011~1101~0 &                              &    \\
             & 0011~0111~1101~0 &                              &    \\
             & 0011~0110~1101~0 &                              &    \\
             & 0001~1000        &                              &    \\
      LOP3   & 0011~11          & 010~0001~0010  & ~010~0001~0010  \\
             & 0101~1011~1110~0 &                              &    \\
             & 0000~001         &                              &    \\
      LOP    & 0101~1100~0100~0 &                              & -  \\
             & 0100~1100~0100~0 &                              &    \\
      \multicolumn{4}{c}{Integer Instructions (continued)} \\[4pt]
      Instruction &  \multicolumn{1}{c}{Pascal} & \multicolumn{1}{c}{Volta} & \multicolumn{1}{c}{Turing} \\
      \midrule
             & 0011~1001~0100~0 &                              &    \\
             & 0011~1000~0100~0 &                              &    \\
      POPC   & 0101~1100~0000~1 & 011~0000~1001  & ~011~0000~1001  \\
             & 0100~1100~0000~1 &                              &    \\
             & 0011~1001~0000~1 &                              &    \\
             & 0011~1000~0000~1 &                              &    \\
      SHF    & 0101~1011~1111~1 &  10~0001~1001  & ~10~0001~1001  \\
             & 0011~0111~1111~1 &                              &    \\
             & 0011~1000~1111~1 &                              &    \\
             & 0011~1001~1111~1 &                              &    \\
             & 0011~0110~1111~1 &                              &    \\
             & 0101~1100~1111~1 &                              &    \\
      SHL    & 0101~1100~0100~1 & *                            & * \\ 
             & 0011~1000~0100~1 &                              &    \\
             & 0011~1001~0100~1 &                              &    \\
             & 0100~1100~0100~1 &                              &    \\
      SHR    & 0101~1100~0010~1 & *                            & * \\ 
             & 0011~1000~0010~1 &                              &    \\
             & 0011~1001~0010~1 &                              &    \\
             & 0100~1100~0010~1 &                              &    \\
      XMAD   & 0101~1011~00            & --   &  -- \\
             & 0100~111                &      &    \\
             & 0101~0001~0             &      &    \\
             & 0011~0111~00            &      &    \\
             & 0011~0110~00            &      &    \\
    VABSDIFF & --  & 10~0001~0100   & ~10~0001~0100  \\
    VABSDIFF4& --  & 10~0001~0101   & ~10~0001~0101  \\
      BREV   & --  & 011~0000~0001  & ~011~0000~0001  \\
      IABS   & --  & 010~0001~0011  & ~010~0001~0011  \\
      IDP    & --  & 010~0010~0110  & ~010~0010~0110  \\
      QSPC   & --  & 0~0011~1010~1010  & ~0~0011~1010~1010  \\
      BMSK   & --  & 010~0001~1011  & ~010~0001~1011  \\
      \midrule
      & & \\
      & & \\
      \multicolumn{4}{c}{Conversion Instructions} \\[4pt]
      Instruction &  \multicolumn{1}{c}{Pascal} & \multicolumn{1}{c}{Volta} & \multicolumn{1}{c}{Turing} \\
      \midrule
      MOV    & 0101~1100~1001~1 & 010~0000~0010  & ~010~0000~0010  \\
             & 0100~1100~1001~1 &                              &    \\
             & 0011~1001~1001~1 &                              &    \\
             & 0011~1000~1001~1 &                              &    \\
      PRMT   & 0101~1011~1100   &  10~0001~0110  & ~10~0001~0110  \\
             & 0101~0011~1100   &                              &    \\
             & 0100~1011~1100   &                              &    \\
             & 0011~0111~1100   &                              &    \\
             & 0011~0110~1100   &                              &    \\
      SEL    & 0101~1100~1010~0 & 010~0000~0111  & ~010~0000~0111  \\
             & 0011~1000~1010~0 & &    \\
             & 0011~1001~1010~0 & &    \\
             & 0100~1100~1010~0 & &    \\
      SHFL   & 1110~1111~0001~0 & 0~1001~1000~1001  & ~0~1001~1000~1001  \\
      CSET   & 0101~0000~1001~1 & --   &   \\
      CSETP  & 0101~0000~1010~0 & --   &   \\
      \multicolumn{4}{c}{Conversion Instructions (continued)} \\[4pt]
      Instruction &  \multicolumn{1}{c}{Pascal} & \multicolumn{1}{c}{Volta} & \multicolumn{1}{c}{Turing} \\
      \midrule
      PSET   & 0101~0000~1000~1 & -- &   \\
      PSETP  & 0101~0000~1001~0 & -- &   \\
      P2R    & 0101~1100~1110~1 & 010~0000~0011     & 010~0000~0011  \\
             & 0100~1100~1110~1 & & \\
             & 0011~1001~1110~1 & & \\
             & 0011~1000~1110~1 & & \\
      R2P    & 0101~1100~1111~0 & 010~0000~0100     & 010~0000~0100  \\
             & 0100~1100~1111~0 & & \\
             & 0011~1001~1111~0 & & \\
             & 0011~1000~1111~0 & & \\
      GETLMEMBASE & --          & 0~0011~1100~0000  & ~0~0011~1100~0000  \\
      \midrule
             & & \\
             & & \\
      \multicolumn{4}{c}{Load/Store Instructions} \\[4pt]
      Instruction &  \multicolumn{1}{c}{Pascal} & \multicolumn{1}{c}{Volta} & \multicolumn{1}{c}{Turing} \\
      \midrule
      LD     & 100              & 0~1001~1000~0000  & 0~1001~1000~0000 \\
      LDC    & 1110~1111~1001~0 & 0~1011~1000~0010  & 0~1011~1000~0010  \\
      LDG    & 1110~1110~1101~0 & 0~0011~1000~0001  & 0~0011~1000~0001  \\
      LDL    & 1110~1111~0100~0 & 0~1001~1000~0011  & 0~1001~1000~0011  \\
      LDS    & 1110~1111~0100~1 & 0~1001~1000~0100  & 0~1001~1000~0100  \\
      ST     & 101              & 0~0011~1000~0101  & 0~0011~1000~0101  \\
      STG    & 1110~1110~1101~1 & 0~0011~1000~0110  & 0~0011~1000~0110  \\
      STL    & 1110~1111~0101~0 & 0~0011~1000~0111  & 0~0011~1000~0111  \\
      STS    & 1110~1111~0101~1 & 0~0011~1000~1000  & 0~0011~1000~1000  \\
      ATOM   & 1110~1101        & 0~0011~1000~1010  & 0~0011~1000~1010  \\
             & 1110~1110~011    & 0~0011~1000~1011  & 0~0011~1000~1011  \\
             & 1110~1110~1111   &                                        &    \\
      ATOMS  & 1110~1100        & 0~0011~1000~1100  & 0~0011~1000~1100  \\
             & 1110~1110~00     & 0~0011~1000~1101  & 0~0011~1000~1101   \\
             & 1110~1110~010    &                                        &    \\
      ATOMG  & --               & 0~0011~1010~1000  & ~0~0011~1010~1000  \\
             &                  & 0~0011~1010~1001  & ~0~0011~1010~1001  \\
      RED    & 1110~1011~1111~1 & 0~1001~1000~1110  & ~0~1001~1000~1110  \\
      CCTL   & 1110~1111~0111   & 0~1001~1000~1111  & ~0~1001~1000~1111  \\
      MEMBAR & 1110~1111~1001~1 & 0~1001~1001~0010  & ~0~1001~1001~0010  \\
      ERRBAR & --               & 0~1001~1010~1011  & ~0~1001~1010~1011  \\

      CCTL   & --               &                                        & 0~1001~1000~1111  \\
      CCTLT  & 1110~1011~1111~0 &                                        &   \\
      CCTLL  &                  & 0~1001~1001~0000  &   \\
      MATCH  & --               & 0~0011~1010~0001  & ~0~0011~1010~0001  \\

      \midrule
             & & & \\
             & & & \\
      \multicolumn{4}{c}{Control Instructions} \\[4pt]
      Instruction &  \multicolumn{1}{c}{Pascal} & \multicolumn{1}{c}{Volta} & \multicolumn{1}{c}{Turing} \\
      \midrule
      BRA    & 1110~0010~0100          & 0~1001~0100~0111  & 0~1001~0100~0111  \\
      BRX    & 1110~0010~0101          & 0~1001~0100~1001  & 0~1001~0100~1001  \\
      JMP    & 1110~0010~0001          & 0~1001~0100~1010  & 0~1001~0100~1010  \\
      JMX    & 1110~0010~0000          & 0~1001~0100~1100  & 0~1001~0100~1100  \\
      SSY    & 1110~0010~1001          &                   &   \\
      SYNC   & 1111~0000~1111~1        & --                &   \\

      \multicolumn{4}{c}{Control Instructions (continued)} \\[4pt]
      Instruction &  \multicolumn{1}{c}{Pascal} & \multicolumn{1}{c}{Volta} & \multicolumn{1}{c}{Turing} \\
      \midrule
      BSYNC  & --                      & 0~1001~0100~0001  & 0~1001~0100~0001  \\
      WARPSYNC & --                    & 011~0100~1000     & 011~0100~1000  \\
      CAL    & 1110~0010~0110          & --                &  \\
      CALL   & --                      & 011~0100~0011     & ~011~0100~0011  \\
             &                         & 0~1001~0100~0100  & ~0~1001~0100~0100   \\
      JCAL   & 1110~0010~0010          & --                &   \\
      PRET   & 1110~0010~0111          & --                &   \\
      RET    & 1110~0011~0010          & 0~1001~0101~0000  & ~0~1001~0101~0000  \\
      BRK    & 1110~0011~0100          & --                &   \\
      PBK    & 1110~0010~1010          & --                &   \\
      CONT   & 1110~0011~0101          & --                &   \\
      PCNT   & 1110~0010~1011          & --                &   \\
      EXIT   & 1110~0011~0000          & 0~1001~0100~1101  & 0~1001~0100~1101  \\
      PEXIT  & 1110~0010~0011          & -- &   \\
      BPT    & 1110~0011~1010          &    &   \\
      BMOV   & --                      & 0~0011~0101~0101  & 0~0011~0101~0101  \\
             &                         & 011~0101~0110     & 011~0101~0110   \\
             &                         & 011~0101~0111     & 011~0101~0111   \\
      YIELD  & --                      & 0~1001~0100~0110  & 0~1001~0100~0110  \\
      RTT    & --                      & 0~1001~0100~1111  & 0~1001~0100~1111  \\
      KILL   & --                      & 0~1001~0101~1011  & 0~1001~0101~1011  \\
      RPCMOV & --                      & 011~0101~0010     & 011~0101~0010  \\
             &                         & 0~0011~0101~0011  & 0~0011~0101~0011   \\
             &                         &                   & 0~1001~0101~0100   \\
      IDE    & --                      & 0~1001~0101~0001  & 0~1001~0101~0001  \\
      PMTRIG & --                      & 0~1000~0000~0001  & 0~1000~0000~0001  \\
      BREAK  & --                      & 0~1001~0100~0010  & 0~1001~0100~0010  \\
      BSSY   & --                      & 0~1001~0100~0101  & 0~1001~0100~0101  \\
      NANOSLEEP & --                   & 011~0101~1101     & 011~0101~1101 \\
      NANOTRAP  & --                   & 011~0101~1010     & 011~0101~1010 \\
      \midrule
             & & \\
             & & \\
      \multicolumn{4}{c}{Other Instructions} \\[4pt]
      Instruction &  \multicolumn{1}{c}{Pascal} & \multicolumn{1}{c}{Volta} & \multicolumn{1}{c}{Turing} \\
      \midrule
      NOP    & 0101~0000~1011~0 & 0~1001~0001~1000  & ~0~1001~0001~1000  \\
      CS2R   & 0101~0000~1100~1 & 0~1000~0000~0101  & ~0~1000~0000~0101  \\
      S2R    & 1111~0000~1100~1 & 0~1001~0001~1001  & ~0~1001~0001~1001  \\
      B2R    & 1111~0000~1011~1 & 0~0011~0001~1100  & ~0~0011~0001~1100  \\
      BAR    & 1110~0010~0100   & 011~0001~1101     & ~011~0001~1101  \\
      R2B    & 1111~0000~1100~0 & 0~0011~0001~1110  & ~0~0011~0001~1110  \\
      VOTE   & 0101~0000~1101~1 & 0~1000~0000~0110  & ~0~1000~0000~0110  \\
             & 0101~0000~1110~0 &                   &    \\
      TMML   & -- & 0~1011~0110~1001  & 0~1011~0110~1001  \\
      TXD    & -- & 0~1011~0110~1100  & 0~1011~0110~1100  \\
      SGXT   & -- & 010~0001~1010     &    010~0001~1010  \\
      AL2P   & -- & --                & 0~1001~0010~0000 \\
      CSMTEST& -- & 0~1000~0000~1101  & 0~1000~0000~1101  \\
      DEPBAR & -- & 0~1001~0001~1010  & 0~1001~0001~1010 \\
      IPA    & -- & --                & 0~0011~0010~0110 \\
      ISBERD & -- & --                & 0~1001~0010~0011 \\
      LEPC   & -- & 0~0011~0100~1110  & 0~0011~0100~1110 \\
      OUT    & -- & --                & 0~0011~0010~0100 \\
      \multicolumn{4}{c}{Other Instructions (continued)} \\[4pt]
      Name        & Pascal & Volta         & Turing             \\
      \midrule
      PIXLD       & -- & --                & 0~1001~0010~0101 \\
      PLOP3       & -- & 0~1000~0001~1100  & 0~1000~0001~1100 \\
      SETCTAID    & -- & 0~0011~0001~1111  & 0~0011~0001~1111 \\
      SETLMEMBASE & -- & 0~0011~1100~0001  & 0~0011~1100~0001 \\
      \bottomrule
  \end{longtable}
}


\bibliographystyle{IEEEtran}
\raggedright
\bibliography{tech_report}

\begin{thebibliography}{10}
\providecommand{\url}[1]{#1}
\csname url@samestyle\endcsname
\providecommand{\newblock}{\relax}
\providecommand{\bibinfo}[2]{#2}
\providecommand{\BIBentrySTDinterwordspacing}{\spaceskip=0pt\relax}
\providecommand{\BIBentryALTinterwordstretchfactor}{4}
\providecommand{\BIBentryALTinterwordspacing}{\spaceskip=\fontdimen2\font plus
\BIBentryALTinterwordstretchfactor\fontdimen3\font minus
  \fontdimen4\font\relax}
\providecommand{\BIBforeignlanguage}[2]{{%
\expandafter\ifx\csname l@#1\endcsname\relax
\typeout{** WARNING: IEEEtran.bst: No hyphenation pattern has been}%
\typeout{** loaded for the language `#1'. Using the pattern for}%
\typeout{** the default language instead.}%
\else
\language=\csname l@#1\endcsname
\fi
#2}}
\providecommand{\BIBdecl}{\relax}
\BIBdecl

\bibitem{zhe2018gtc}
\BIBentryALTinterwordspacing
Z.~Jia, M.~Maggioni, B.~Staiger, and D.~P. Scarpazza, ``Dissecting the {NVidia}
  {Volta} {GPU} architecture via microbenchmarking,'' in \emph{2018 GPU
  Technology Conference}, 2018. [Online]. Available:
  \url{http://on-demand.gputechconf.com/gtc/2018/presentation/s8122-dissecting-the-volta-gpu-architecture-through-microbenchmarking.pdf}
\BIBentrySTDinterwordspacing

\bibitem{zhe2018}
\BIBentryALTinterwordspacing
------, ``Dissecting the {NVidia} {Volta} {GPU} architecture via
  microbenchmarking,'' 2018. [Online]. Available:
  \url{https://arxiv.org/abs/1804.06826}
\BIBentrySTDinterwordspacing

\bibitem{tu104}
\emph{{NVidia} Turing GPU Architecture}.\hskip 1em plus 0.5em minus 0.4em\relax
  NVIDIA Corporation, 2018.

\bibitem{asfermi}
\BIBentryALTinterwordspacing
Y.~Hou, ``Asfermi,'' 2011. [Online]. Available:
  \url{https://github.com/hyqneuron/asfermi}
\BIBentrySTDinterwordspacing

\bibitem{askepler}
C.~Wang, Z.~Jia, and K.~Chen, ``Tuning performance on {Kepler} {GPUs}: An
  introduction to {Kepler} assembler and its usage in {CNN} optimization,'' in
  \emph{GPU Technology Conference Presentation}, vol. 6173, 2015.

\bibitem{nervana}
\BIBentryALTinterwordspacing
S.~Gray, ``maxas,'' 2016. [Online]. Available:
  \url{https://github.com/NervanaSystems/maxas/wiki/Control-Codes}
\BIBentrySTDinterwordspacing

\bibitem{wong2010}
H.~Wong, M.~M. Papadopoulou, M.~Sadooghi-Alvandi, and A.~Moshovos,
  ``Demystifying {GPU} microarchitecture through microbenchmarking,'' in
  \emph{2010 IEEE International Symposium on Performance Analysis of Systems
  Software (ISPASS)}, March 2010, pp. 235--246.

\bibitem{zhang2017}
X.~Zhang, G.~Tan, S.~Xue, J.~Li, K.~Zhou, and M.~Chen, ``Understanding the
  {GPU} microarchitecture to achieve bare-metal performance tuning,'' in
  \emph{Proceedings of the 22nd ACM SIGPLAN Symposium on Principles and
  Practice of Parallel Programming}, ser. PPoPP '17.\hskip 1em plus 0.5em minus
  0.4em\relax New York, NY, USA: ACM, 2017, pp. 31--43.

\bibitem{mei2017}
X.~Mei and X.~Chu, ``Dissecting {GPU} memory hierarchy through
  microbenchmarking,'' \emph{IEEE Transactions on Parallel and Distributed
  Systems}, vol.~28, no.~1, pp. 72--86, Jan 2017.

\bibitem{gv100}
\emph{{NVidia} {Tesla} V100 GPU Architecture, The World's Most Advanced Data
  Center GPU}.\hskip 1em plus 0.5em minus 0.4em\relax NVIDIA Corporation, 2017.

\bibitem{cuobj}
\BIBentryALTinterwordspacing
``{NVidia} cuobjdump and nvdisasm,'' 2016. [Online]. Available:
  \url{https://docs.nvidia.com/cuda/cuda-binary-utilities/}
\BIBentrySTDinterwordspacing

\end{thebibliography}

\newpage
\tableofcontents

\chapter*{The Authors}
\newlength{\picsize}
\newlength{\biosize}
\setlength{\picsize}{0.8in}
\setlength{\biosize}{\textwidth-\picsize}

\begin{tabular}{p{\picsize}p{\biosize}}
  \raisebox{-\height}{\includegraphics[width=\picsize]{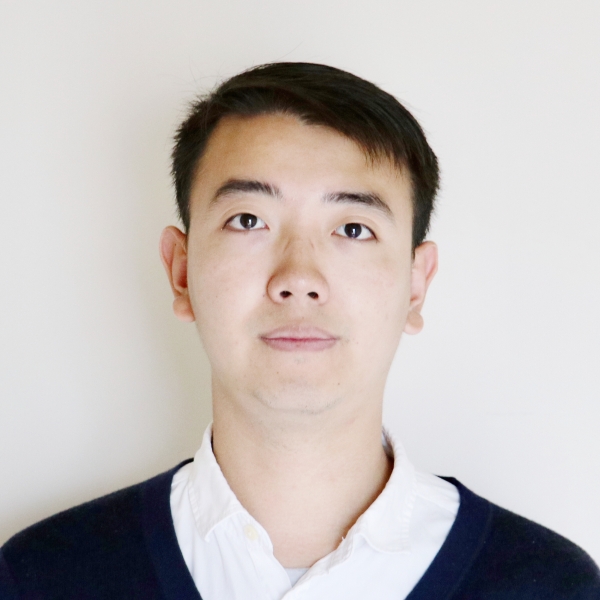}}
  & \sffamily\textbf{Zhe Jia} is a senior R\&D engineer with the
  High-Performance Computing group at Citadel.  Prior to this
  position, he was a senior R\&D engineer with Alicloud, and a
  software engineering intern at Microsoft, Beijing.  He received his
  B.S. degree in Physics and his M.S. degree in Meteorology from
  Peking University, China.  His interests include the performance
  optimization of deep learning, numerical modeling, and atmospheric
  simulation workloads. \\
  & \\
  \raisebox{-\totalheight}{\includegraphics[width=\picsize]{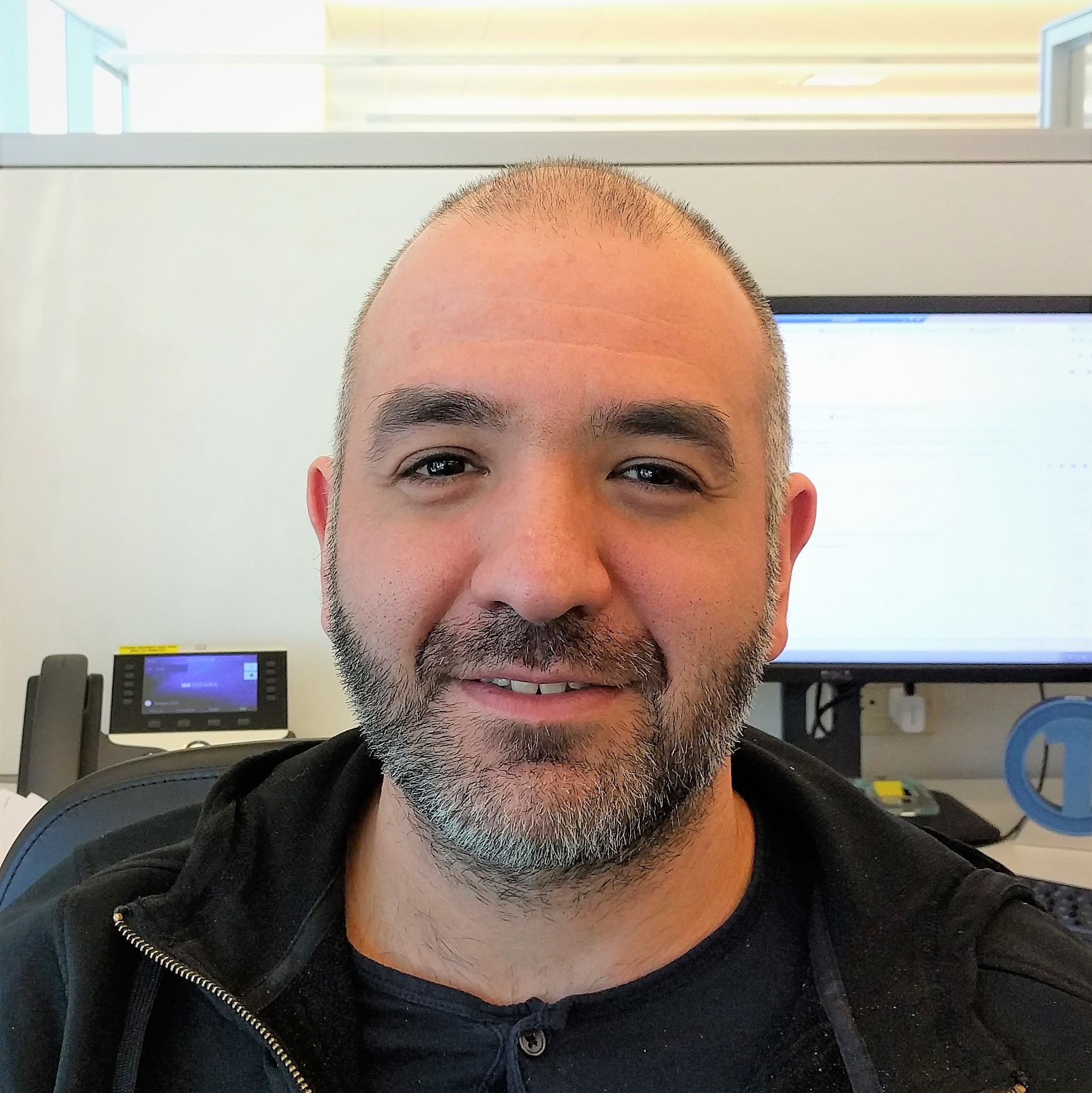}}
  & \sffamily\textbf{Marco Maggioni} Marco is a senior R\&D engineer with the
  High-Performance Computing group at Citadel, Chicago.  He received
  his Ph.D. in Computer Science from the University of Illinois at
  Chicago, where he focused on sparse linear algebra and convex
  optimization on GPUs.  His research established records for the fastest
  sparse matrix-vector multiplication GPU kernel in research and
  industry. \\
  & \\
  \raisebox{-\totalheight}{\includegraphics[width=\picsize]{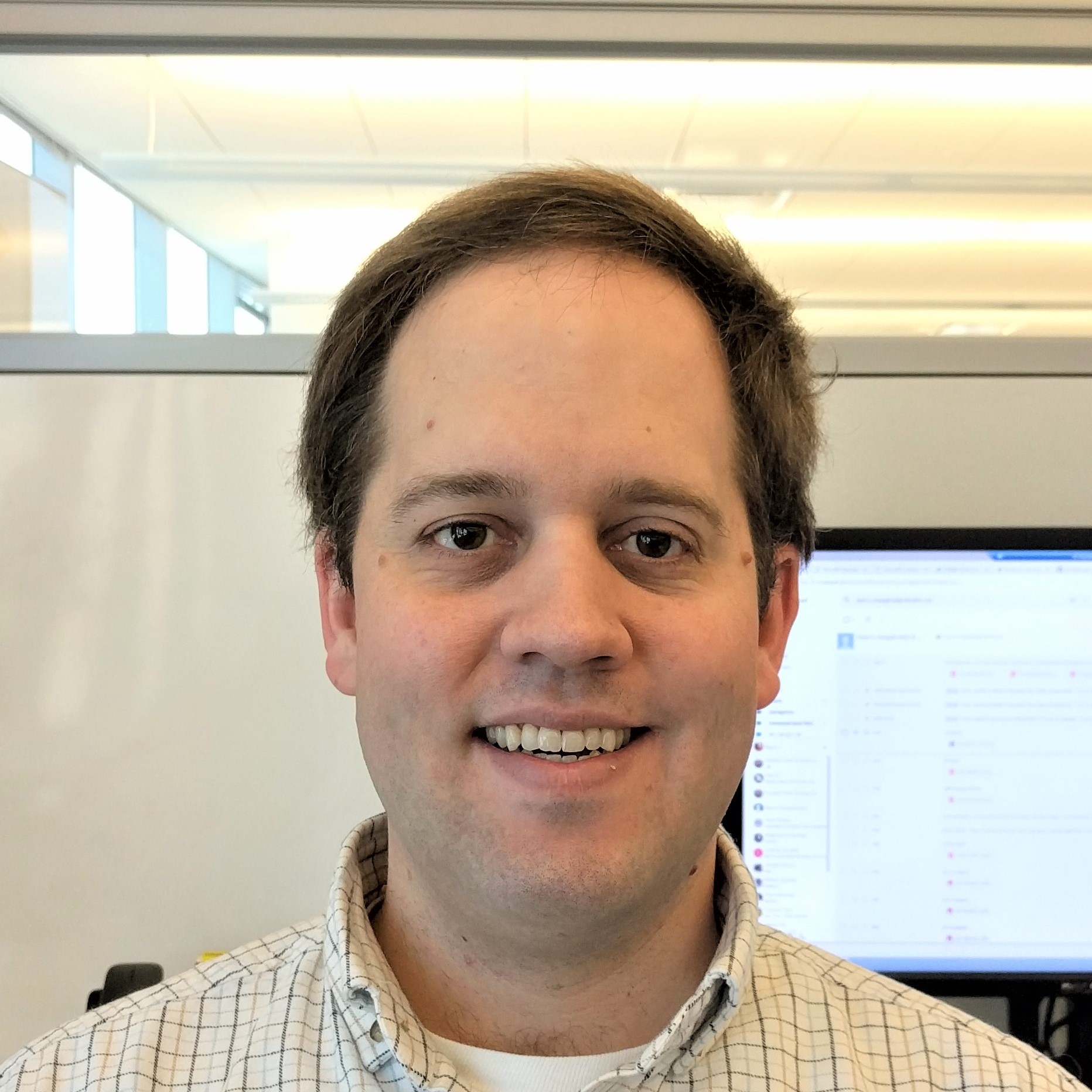}}
  & \sffamily\textbf{Jeffrey Smith} is a senior software engineer with the
  Market-Making Trading Platform R\&D team at Citadel Securities. He
  received his B.S. in Computer Science from Rose-Hulman Institute of
  Technology. His interests include hardware architectural analysis,
  latency-focused systems acceleration and design, and software
  framework creation for thread- and network-level task
  parallelization.\\
  & \\
  \raisebox{-\totalheight}{\includegraphics[width=\picsize]{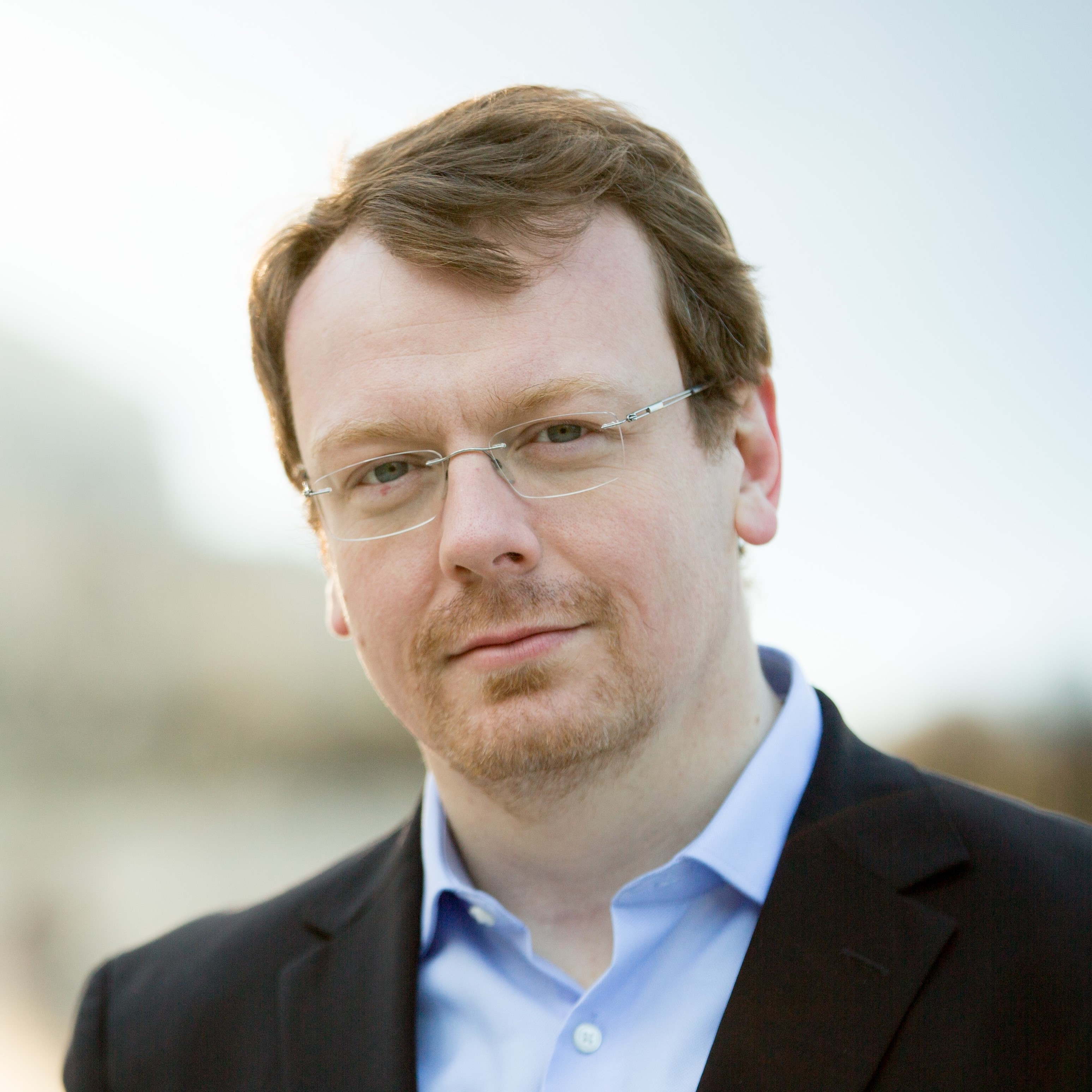}}
  & \sffamily\textbf{Daniele Paolo Scarpazza} Daniele leads the
  High-Performance Computing group at Citadel, Chicago.  Prior to this
  position, he was a Research Scientist with D. E. Shaw Research, a
  Research Staff Member with the IBM T. J. Watson Research Center, and
  a Post-Doc with the Pacific Northwest National Laboratory.  He
  received his Ph.D. in Information Engineering from Politecnico di
  Milano, Italy.  He is the co-recipient of a Gordon Bell prize.  He
  focuses on quantitative performance analysis and optimization of
  diverse algorithms on parallel architectures.
\end{tabular}

\end{document}